\documentclass[sigconf]{acmart}

\usepackage{color,xcolor}
\usepackage{graphicx}

\usepackage{array}
\usepackage{booktabs}
\usepackage{colortbl}
\usepackage{float,wrapfig}
\usepackage{multirow}
\usepackage{subcaption} %
\usepackage[toc,page]{appendix}

\usepackage{amsfonts,amsthm}
\usepackage{bm}
\usepackage{nicefrac}
\usepackage{microtype}
\usepackage{pifont}

\usepackage{fancyhdr}
\usepackage{setspace}
\usepackage{soul}
\usepackage{xspace}

\usepackage{algorithm, algorithmic}

\definecolor{hlcolor}{RGB}{255,252,228}
\newcolumntype{L}[1]{>{\raggedright\let\newline\\\arraybackslash\hspace{0pt}}m{#1}}
\newcolumntype{C}[1]{>{\centering\let\newline\\\arraybackslash\hspace{0pt}}m{#1}}
\newcolumntype{R}[1]{>{\raggedleft\let\newline\\\arraybackslash\hspace{0pt}}m{#1}}

\newcommand{\sect}[1]{Section~\ref{#1}}
\newcommand{\eqn}[1]{Equation~\ref{#1}}
\newcommand{\fig}[1]{Figure~\ref{#1}}
\newcommand{\tab}[1]{Table~\ref{#1}}

\newcommand{\ignore}[1]{}

\makeatletter
\DeclareRobustCommand\onedot{\futurelet\@let@token\@onedot}
\def\@onedot{\ifx\@let@token.\else.\null\fi\xspace}

\def\eg{\emph{e.g}\onedot} 
\def\ie{\emph{i.e}\onedot} 
 
 \def\vs{\emph{vs}\onedot}
 
\def\etal{\emph{et al}\onedot}
\makeatother

\newcommand{\OffsetSet}[2]{\ensuremath{\bm{\Delta}^{#1}(#2)}}
\definecolor{citecolor}{HTML}{64287e}
\definecolor{mydarkblue}{rgb}{0,0.08,1}
\definecolor{mydarkgreen}{rgb}{0.02,0.6,0.02}
\definecolor{mydarkred}{rgb}{0.8,0.02,0.02}
\definecolor{mydarkorange}{rgb}{0.40,0.2,0.02}
\definecolor{mypurple}{RGB}{111,0,255}
\definecolor{myred}{rgb}{1.0,0.0,0.0}
\definecolor{mygold}{rgb}{0.75,0.6,0.12}
\definecolor{mydarkgray}{rgb}{0.66, 0.66, 0.66}
\definecolor{linkcolor}{HTML}{000000}%

\def\system{TorchSparse++\xspace}
\def\systemv1{TorchSparse\xspace}
\def\SparseConv{SparseConv\xspace}
\def\ME{MinkowskiEngine\xspace}
\def\SpConv{SpConv\xspace}

\usepackage{listings}
\usepackage{xcolor}

\definecolor{codegreen}{rgb}{0,0.6,0}
\definecolor{codegray}{rgb}{0.5,0.5,0.5}
\definecolor{codepurple}{rgb}{0.58,0,0.82}
\definecolor{backcolour}{rgb}{0.95,0.95,0.92}

\lstdefinestyle{mystyle}{
    backgroundcolor=\color{backcolour},   
    commentstyle=\color{codegreen},
    keywordstyle=\color{magenta},
    numberstyle=\tiny\color{codegray},
    stringstyle=\color{codepurple},
    basicstyle=\ttfamily\footnotesize,
    breakatwhitespace=false,         
    breaklines=true,                 
    captionpos=b,                    
    keepspaces=true,                 
    numbers=left,                    
    numbersep=5pt,                  
    showspaces=false,                
    showstringspaces=false,
    showtabs=false,                  
    tabsize=2
}

\AtBeginDocument{%
  }

\setcopyright{acmcopyright}
\copyrightyear{2023} 
\acmYear{2023} 
\setcopyright{rightsretained} 
\acmConference[MICRO '23]{56th Annual IEEE/ACM International Symposium on Microarchitecture}{October 28-November 1, 2023}{Toronto, ON, Canada}
\acmBooktitle{56th Annual IEEE/ACM International Symposium on Microarchitecture (MICRO '23), October 28-November 1, 2023, Toronto, ON, Canada}
\acmDOI{10.1145/3613424.3614303}
\acmISBN{979-8-4007-0329-4/23/10}

\begin{document}

\title[TorchSparse++]{TorchSparse++: Efficient Training and Inference Framework for Sparse Convolution on GPUs}

\author{Haotian Tang}
\authornote{Both authors contributed equally to this research.}
\affiliation{%
  \institution{MIT}
  \city{Cambridge}
  \state{MA}
  \country{USA}
}
\email{kentang@mit.edu}

\author{Shang Yang}
\authornotemark[1]
\affiliation{%
  \institution{MIT, Tsinghua University}
  \city{Cambridge}
  \state{MA}
  \country{USA}}
\email{shangy@mit.edu}

\author{Zhijian Liu}
\affiliation{%
  \institution{MIT}
  \city{Cambridge}
  \state{MA}
  \country{USA}}
\email{zhijian@mit.edu}

\author{Ke Hong}
\affiliation{%
  \institution{Tsinghua University}
  \city{Beijing}
  \country{China}}
\email{hjkl1379991@126.com}

\author{Zhongming Yu}
\affiliation{%
  \institution{UCSD}
  \city{San Diego}
  \state{CA}
  \country{USA}}
\email{zhy025@ucsd.edu}

\author{Xiuyu Li}
\affiliation{%
  \institution{UC Berkeley}
  \city{Berkeley}
  \state{CA}
  \country{USA}}
\email{xiuyu@berkeley.edu}

\author{Guohao Dai}
\affiliation{%
  \institution{Shanghai Jiao Tong University}
  \city{Shanghai}
  \country{China}}
\email{daiguohao@sjtu.edu.cn}

\author{Yu Wang}
\affiliation{%
  \institution{Tsinghua University}
  \city{Beijing}
  \country{China}}
\email{yu-wang@tsinghua.edu.cn}

\author{Song Han}
\affiliation{%
  \institution{MIT, NVIDIA}
  \city{Cambridge}
  \state{MA}
  \country{USA}
}
\email{songhan@mit.edu}

\renewcommand{\shortauthors}{Tang and Yang et al.}

\begin{abstract}
Sparse convolution plays a pivotal role in emerging workloads, including point cloud processing in AR/VR, autonomous driving, and graph understanding in recommendation systems. Since the computation pattern is sparse and irregular, specialized high-performance kernels are required. Existing GPU libraries offer two dataflow types for sparse convolution. The gather-GEMM-scatter dataflow is easy to implement but not optimal in performance, while the dataflows with overlapped computation and memory access (\eg implicit GEMM) are highly performant but have very high engineering costs. In this paper, we introduce \textbf{\system}, a new GPU library that achieves the best of both worlds. We create a highly efficient Sparse Kernel Generator that generates performant sparse convolution kernels at less than one-tenth of the engineering cost of the current state-of-the-art system. On top of this, we design the Sparse Autotuner, which extends the design space of existing sparse convolution libraries and searches for the best dataflow configurations for training and inference workloads. Consequently, \system achieves \textbf{2.9$\times$}, \textbf{3.3$\times$}, \textbf{2.2$\times$} and \textbf{1.7$\times$} measured end-to-end speedup on an NVIDIA A100 GPU over state-of-the-art MinkowskiEngine, SpConv 1.2, TorchSparse and SpConv v2 in inference; and is \textbf{1.2-1.3$\times$} faster than SpConv v2 in mixed precision training across seven representative autonomous driving benchmarks. It also seamlessly supports graph convolutions, achieving \textbf{2.6-7.6$\times$} faster inference speed compared with state-of-the-art graph deep learning libraries. Our code is publicly released at \url{https://github.com/mit-han-lab/torchsparse}. %

\end{abstract}



\maketitle

\section{Introduction}

Sparse convolution~\cite{graham20183d,choy20194d} plays a crucial role in a variety of cutting-edge applications, including augmented/virtual reality (AR/VR), autonomous driving, and recommendation systems. For instance, in advanced driver assistance systems (ADAS) and autonomous driving technology, data is collected from 3D sensors in the form of 3D point clouds. These point clouds often exhibit an exceptionally high spatial sparsity, with up to 99.99\% spatial sparsity. In such cases, employing dense 3D convolutions for point cloud processing becomes inefficient. Likewise, social media graphs, like those found on platforms such as Twitter, exhibit even greater sparsity. As an illustration, the adjacency matrix of Twitter's social graph contains only a minuscule fraction, approximately 0.000214\%, of the possible connections~\cite{zhang2020sparch}. Therefore, there is a urgent need for efficient inference and training system for these sparse workloads.

\begin{figure}
    \centering
    \includegraphics[width=\linewidth]{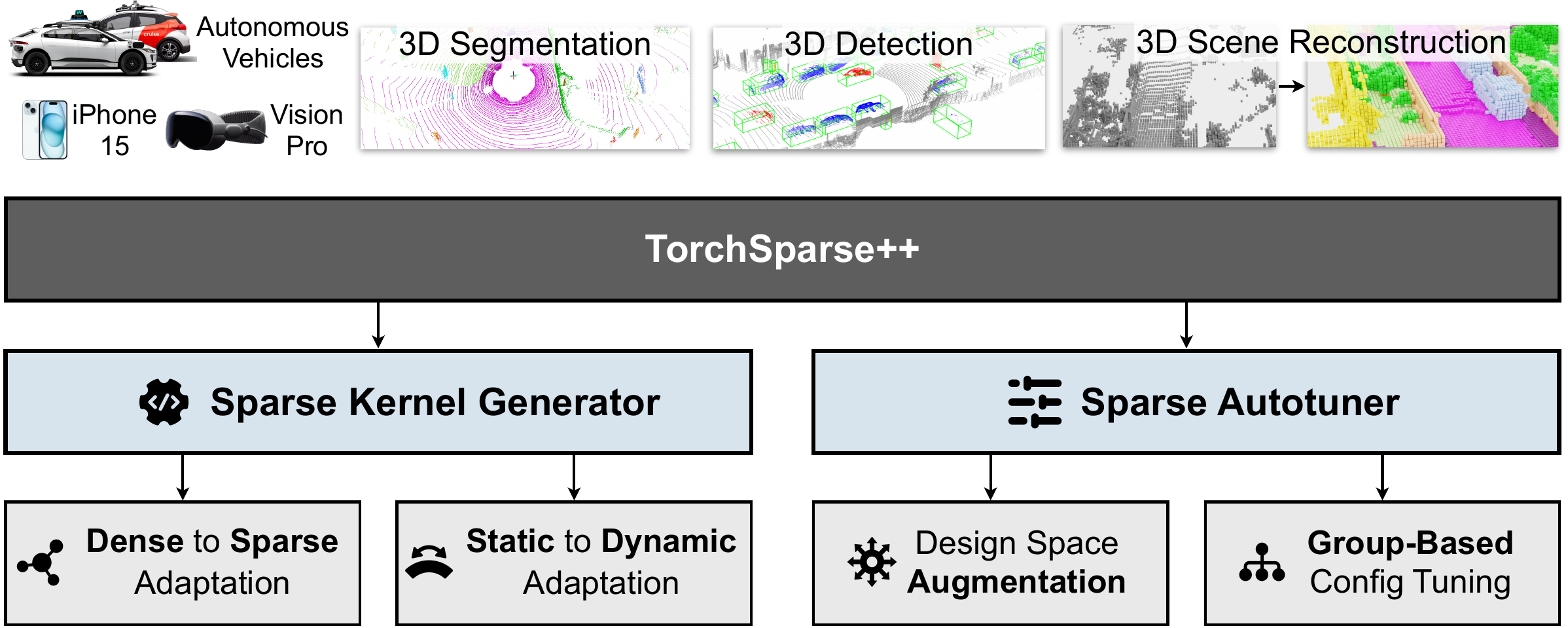}
    \caption{\system is a high-performance GPU library that provides highly-optimized dataflows for convolution on point clouds through a \textbf{Sparse Kernel Generator}. It further selects the optimal dataflow for each layer using the \textbf{Sparse Autotuner}, delivering up to 1.3-1.7$\times$ geomean inference and training speedups than the previous state-of-the-art system. Figures source:~\cite{behley2019semantickitti,cheng2020s3cnet,qi2021offboard}, Waymo, Cruise, Apple.}
    \label{fig:intro:teaser}
\end{figure}

Sparse convolution modifies the definition of regular convolution by only performing computation at a sparse set of output locations rather than the entire feature map. It is arguably the most important building block for almost all state-of-the-art 3D perception models (\eg 3D semantic segmentation~\cite{tang2020searching,liu2021pvnas,cheng2021af2s3net}, 3D object detection~\cite{yan2018second,yin2021center,zhou2022centerformer,chen2022mppnet,ye2022lidarmultinet,chen2022scaling,ge2021afdet,bai2022transfusion}, 3D reconstruction~\cite{cheng2020s3cnet}, multi-sensor fusion~\cite{chen2022focal,liu2023bevfusion,li2022uvtr}, end-to-end navigation~\cite{liu2021efficient}). It also exhibits similar computation pattern to (relational) graph convolutions~\cite{hamilton2017inductive,schlichtkrull2018modeling}. Despite achieving dominant performance, the sparse and irregular nature of sparse convolution makes it harder to be processed on GPUs and there is no vendor library support. Dedicated libraries~\cite{graham20183d,yan2018second,yan2022spconv,tang2022torchsparse,hong2023pcengine} with specialized high-performance kernels or even specialized hardware accelerators~\cite{feng2020mesorasi,lin2021pointacc,feng2022crescent} are required for sparse convolution. As a result, many industrial driving assistance solutions still prefer pillar-based models~\cite{lang2019pointpillars}, which flatten LiDAR points onto the BEV space and process them with a 2D CNN. These approaches cannot take full advantage of 3D geometry from LiDAR and tend to have much worse accuracy. 

Several pioneering implementations of sparse convolution have adopted different dataflows for this operator. For instance, SparseConvNet~\cite{graham20183d} and SpConv v1~\cite{yan2018second} use the vanilla gather-GEMM-scatter dataflow. It was improved by TorchSparse~\cite{tang2022torchsparse} that optimizes the gather-scatter paradigm through fusing memory operations and grouping computations adaptively into batches to improve device utilization. Dataflows based on gather-scatter can be implemented using vendor libraries with relative ease. However, they are fundamentally restricted in performance due to the inability to overlap memory access and computation.
MinkowskiEngine~\cite{choy20194d} proposes the fetch-on-demand dataflow, which is optimized by PCEngine~\cite{hong2023pcengine}. Recently, SpConv v2~\cite{yan2018second,yan2022spconv} has adapted the implicit GEMM dataflow for dense convolution to the sparse domain, achieving state-of-the-art performance on real-world workloads. Nevertheless, the best representative of these memory-computation overlapped dataflows, implicit GEMM, is extremely hard to implement. The metaprogrammer for \SpConv v2 has more than 40k lines of code, making it hard for the community to further improve upon it.

To address the significant challenge of achieving both ease of implementation and state-of-the-art performance, we present \textit{\system} (\fig{fig:intro:teaser}), a high-performance GPU library that combines the best of both worlds through the \textit{Sparse Kernel Generator} and the \textit{Sparse Autotuner}. Tackling a fundamentally \textit{sparse} and \textit{dynamic} workload, we propose a general method to adapt existing tensor compilers that are optimized for \textit{dense} and \textit{static} workloads, unlocking their potential to generate kernels that can deal with sparsity and variable workload shapes. On top of the generated kernels, we further extend the design space of existing point cloud libraries. We design a Sparse Autotuner to efficiently search for the best dataflow configurations through group-based tuning for a diverse set of workloads within the enlarged design space. The results of our Sparse Autotuner challenged the conventional design wisdom of using amount of computation, DRAM access or even total runtime for computation kernels as the indicator for end-to-end performance. 

As a result, evaluated on seven representative models across three benchmark datasets, \system achieves end-to-end speedups of \textbf{2.9$\times$}, \textbf{3.3$\times$}, \textbf{2.2$\times$}, and \textbf{1.7$\times$} on an NVIDIA A100 GPU, at least \textbf{1.4$\times$}, \textbf{1.8$\times$}, \textbf{2.4$\times$}, \textbf{2.2$\times$} speedup on three other cloud GPUs from Pascal to Ampere architectures in three data precisions over state-of-the-art systems such as MinkowskiEngine, SpConv 1.2, TorchSparse, and SpConv v2. Additionally, in the training phase, our system outperforms the best existing training system, SpConv v2, by \textbf{1.2-1.3$\times$}  under mixed-precision training scenarios. Our system also delivers \textbf{2.6-7.6$\times$} speedup to relational graph convolution workloads over DGL, PyG and Graphiler. Code is publicly released at \url{https://github.com/mit-han-lab/torchsparse}.

\section{Background and Motivation}
\label{sect:background}

Without loss of generality, we use point cloud workloads to illustrate the computation pattern of sparse convolution. A point cloud sparse tensor can be defined as an unordered set of points with features: $\{(\bm{p}_j, \bm{x}_{j})\}$. $\bm{p}_j$ is the quantized coordinates for the $j\textsuperscript{th}$ point in the $D$-dimensional space $\mathbb{Z}^D$. $\bm{x}_j$ is its $C$-dimensional feature vector in $\mathbb{R}^C$. Coordinate quantization is done through $\bm{p} = \lfloor\bm{p}_i^{(\text{raw})} / \bm{v}\rfloor$, where $\bm{v}$ is the voxel size vector. \texttt{Unique} operation is further applied to all quantized coordinates. For example, in CenterPoint~\cite{yin2021center}, the point clouds on Waymo~\cite{sun2020scalability} are quantized using $\bm{v} = $ [0.1m, 0.1m, 0.15m]. This means that we will only keep one point within each 0.1m$\times$0.1m$\times$0.15m grid.

\subsection{Definition of Sparse Convolution}

\begin{figure}
    \centering
    \includegraphics[width=0.9\linewidth]{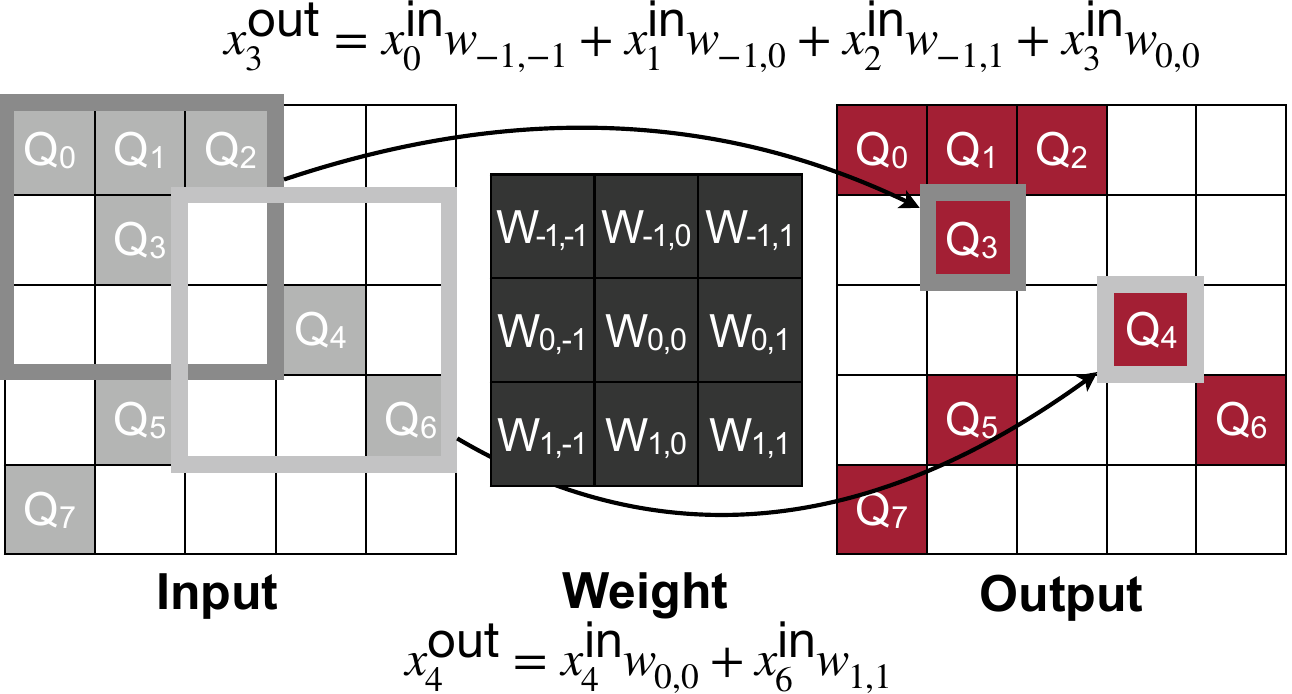}
    \caption{Sparse convolution (\eqn{eqn:sparseconv}) on $\Delta^2(3)$: computation is performed only on \textit{nonzero} inputs.}
    \label{fig:background:workload}
\end{figure}

\begin{figure}
    \centering
    \includegraphics[width=\linewidth]{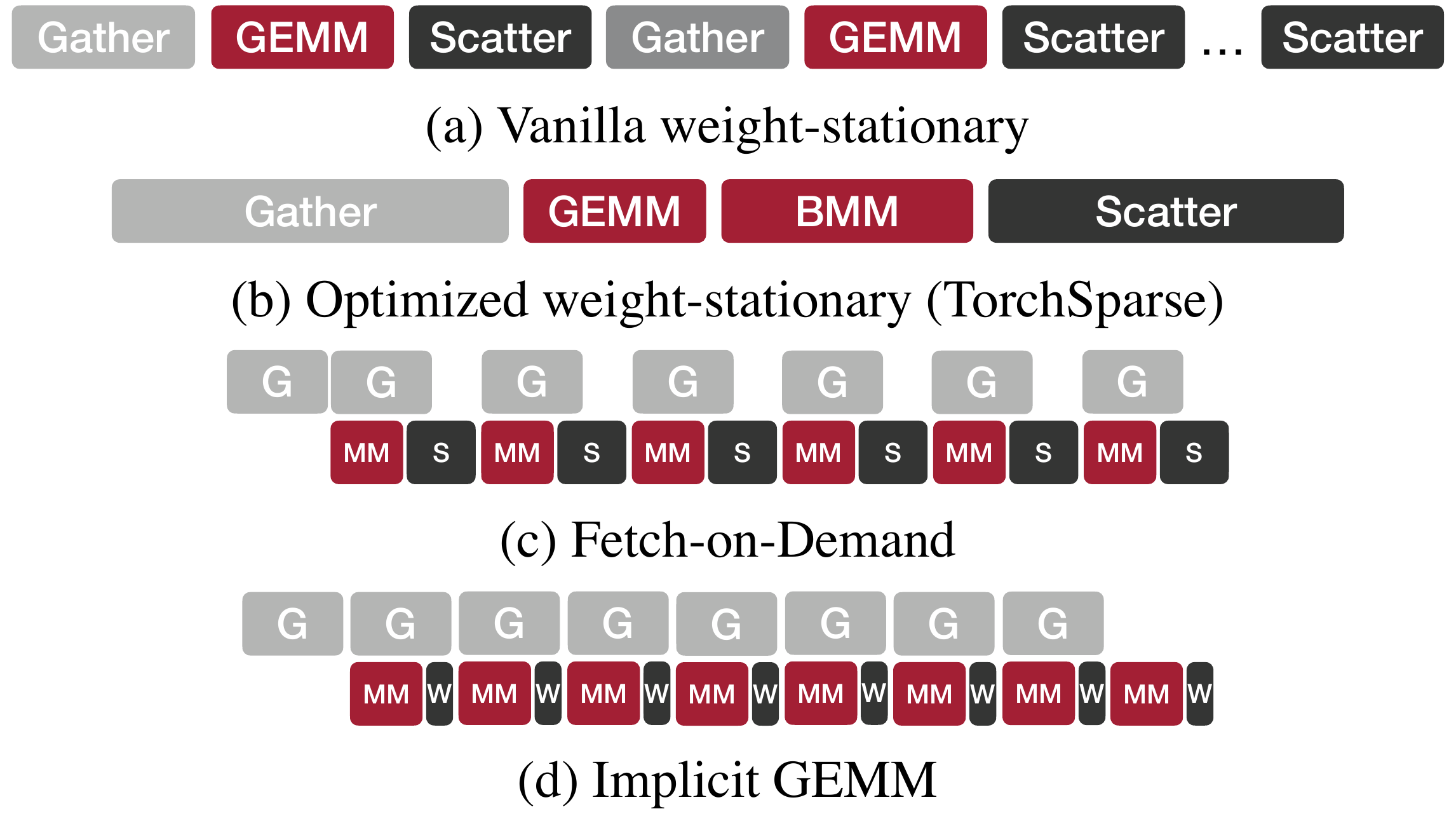}

    \caption{Waterfall diagram for different dataflows for sparse convolution on GPU: weight-stationary dataflows (a, b) are easier to implement and maintain but they do not overlap memory access with computation. Both fetch-on-demand and implicit GEMM dataflows require custom MMA routines but are able to hide the memory access time with pipelining.}
    \label{fig:background:waterfall}
\end{figure}

Following the notations in \cite{tang2022torchsparse}, we define the $D$-dimensional neighborhood with kernel size $K$ as $\Delta^D(K)$ (\eg $\Delta^{2}(5)$ = $\{-2, -1, 0, 1, 2\}^{2}$ and $\Delta^3(3) = \{-1, 0, 1\}^3$). The forward form of sparse convolution (\fig{fig:background:workload}) on the $k\textsuperscript{th}$ output point is defined as:
\begin{equation}\label{eqn:sparseconv}
    \bm{x}_{k}^{\text{out}} = \sum_{\bm{\delta} \in \OffsetSet{D}{K}}\sum_j 1[\bm{p}_j = s \bm{q}_k + \bm{\delta}]~(\bm{x}_{j}^{\text{in}} \cdot \bm{W}_{\bm{\delta}}),
\end{equation}
where $\bm{p}_j\in \bm{P}^{\text{in}}, \bm{q}_k \in \bm{P}^{\text{out}}$, $1[\cdot]$ is a binary indicator, $s$ is the stride and $\bm{W}_{\delta}\in\mathbb{R}^{C\textsubscript{in}\times C\textsubscript{out}}$ corresponds to the weight matrix for kernel offset $\delta\in\Delta^D(K)$. %

\subsection{Sparse Convolution Dataflows on GPUs}

\begin{figure}
    \centering
    \includegraphics[width=0.8\linewidth]{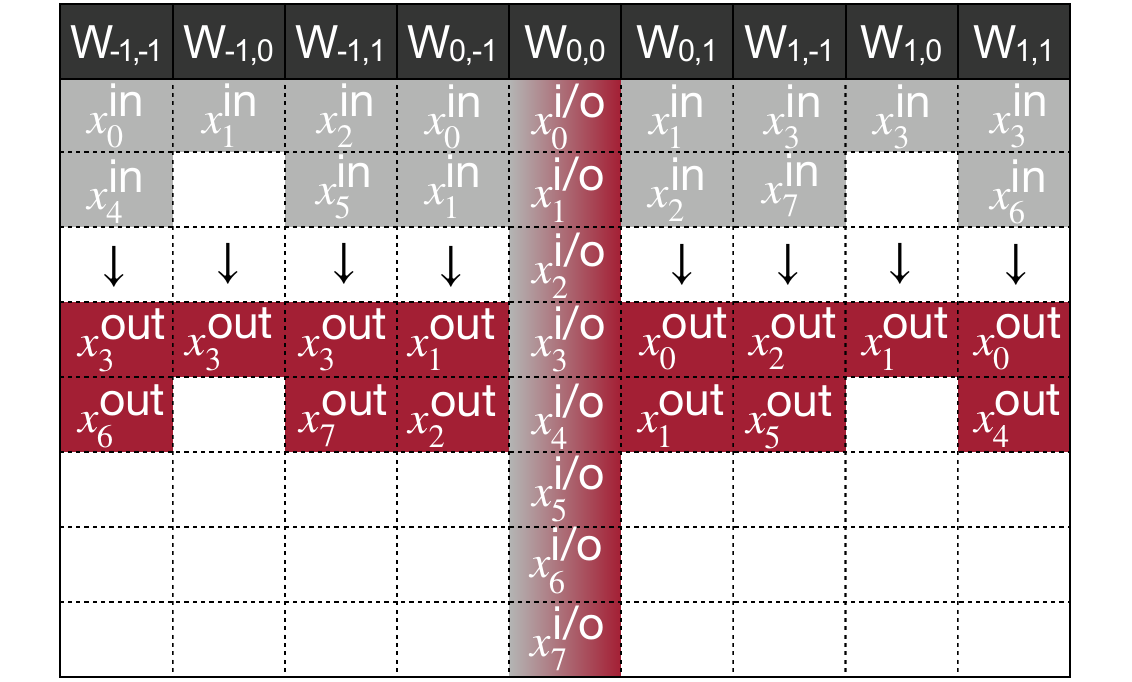}
    \caption{Illustration of the gather-GEMM-scatter dataflow for \fig{fig:background:workload} workload: we first gather input features according to $\mathcal{M}_\delta$ for each weight $\delta$, then perform GEMM or batched GEMM, and finally scatter the results back to output locations given in $\mathcal{M}_\delta$.}
    \label{fig:background:weight_stationary}
\end{figure}

Current implementations of sparse convolution on GPUs can be categorized into three distinct dataflows (\fig{fig:background:waterfall}). The first is the \textit{gather-GEMM-scatter} approach, which is weight-stationary and was inspired by early explicit \texttt{im2col} attempts~\cite{jia2014caffe} for convolution implementation. The second dataflow is the \textit{fetch-on-demand} approach, which is a kernel fusion version of gather-GEMM-scatter. Finally, the \textit{implicit GEMM} approach is an output-stationary alternative inspired by its dense counterpart~\cite{chetlur2014cudnn}.

\subsubsection{Gather-GEMM-Scatter Dataflow}
\label{sect:background:gather_gemm_scatter}

Early sparse convolution implementations utilized a gather-GEMM-scatter dataflow~\cite{graham20183d,yan2018second}. This dataflow is weight-stationary and features an outer host loop over $K^D$ kernel offsets. For each offset $\bm{\delta}\in \Delta^D(K)$, we compute maps $\mathcal{M}_\delta=\{(\bm{p}_j,\bm{q}_k)\lvert \bm{p}_j=s\bm{q}_k+\bm{\delta}\}$, as shown in \fig{fig:background:weight_stationary}. We gather all input features ${\bm{x}^\text{in}_j}$, resulting in a $\lvert\mathcal{M}_\delta\rvert \times C\textsubscript{in}$ matrix in DRAM, and multiply it by weight $\bm{W}_\delta\in\mathbb{R}^{C\textsubscript{in}\times C\textsubscript{out}}$. Finally, we scatter the results back to output positions ${\bm{x}_k^\text{out}}$ according to $\mathcal{M}_\delta$. For example, since $\bm{p}_0=1\times\bm{q}_1+(-1, -1)$ and $\bm{p}_4=1\times\bm{q}_5+(-1, -1)$, $\mathcal{M}_{-1,-1}=\{(\bm{p}_0, \bm{q}_1), (\bm{p}_4, \bm{q}_5)\}$. We gather $\bm{x}_0^\text{in}$ and $\bm{x}_4^\text{in}$, multiply them by $\bm{W}_{-1, -1}$, and scatter the results back to $\bm{x}_1^\text{out}$ and $\bm{x}_5^\text{out}$. A variant of this dataflow~\cite{tang2022torchsparse} aims to reduce both computation and data movement time by fusing and reordering memory accesses and grouping computation for different weights.

Gather-GEMM-scatter is straightforward to implement. Following feature gathering, computation for each offset $\bm{\delta}$ involves a dense matrix multiplication, which can be handled by existing vendor libraries like cuBLAS and cuDNN. Only scatter and gather operations need to be optimized in CUDA. However, this dataflow is fundamentally inefficient due to the lack of overlap between computation and memory access, as illustrated in \fig{fig:background:waterfall}\textcolor{linkcolor}{a,b}. It is thus impossible to hide data orchestration latency with pipelining.

\subsubsection{Fetch-On-Demand Dataflow}

The gather-GEMM-scatter implementation requires three separate CUDA kernel calls in each host loop iteration over $\bm{\delta}$. An alternative fetch-on-demand dataflow~\cite{choy20194d,yan2018second} (named by~\cite{tang2022torchsparse}) merges the gather, matrix multiplication, and scatter kernel calls into a single CUDA kernel. Instead of materializing the $\lvert\mathcal{M}_\delta\rvert\times C_\text{in}$ gather buffer in DRAM, it \underline{fetches} $\{\bm{x}_j^\text{in}\lvert (\bm{p}_j,\bm{q}_k)\in\mathcal{M}_\delta\}$ \underline{on demand} into the L1 shared memory, performs matrix multiplication in the on-chip storage and directly scatters the partial sums (resided in the register file) to corresponding outputs $\{\bm{x}_k^\text{out}\lvert (\bm{p}_j,\bm{q}_k)\in\mathcal{M}_\delta\}$ without first instantiating them in a DRAM scatter buffer. Hong~\etal~\cite{hong2023pcengine} further improve the vanilla fetch-on-demand dataflow by introducing block fusion, where the \textit{sequential} host loop over $\bm{\delta}$ is converted to a \textit{parallel} thread block dimension. As such, the computation of all $\bm{\delta}$s is merged into a single kernel. Similar to gather-GEMM-scatter (without adaptive grouping in~\cite{tang2022torchsparse}), the fetch-on-demand dataflow has \textit{zero} redundant computation. It further overlaps computation with memory access and saves DRAM writes to gather and scatter buffers. 

However, it cannot save any DRAM write to the final output tensor, which means $\frac{\sum_{\bm{\delta}} \lvert\mathcal{M}_{\bm{\delta}}\rvert}{N_\text{out}}\times$ (4$\times$-10$\times$ in real workloads, since each point typically has 4-10 neighbors) larger write-back traffic than the theoretical optimal value. Furthermore, the block-fused fetch-on-demand dataflow~\cite{hong2023pcengine} suffers from write-back contentions between different threads. For example, both $\bm{W}_{-1,0}$ and $\bm{W}_{-1,1}$ in \fig{fig:background:weight_stationary} may attempt to write back to $\bm{x}_3^{\text{out}}$. Therefore, it is necessary to introduce atomic operations to serialize all DRAM writes to the same location. Since gather and scatter operations are now combined into GEMM, the entire computation kernel in the fetch-on-demand dataflow must be implemented in CUDA. This is more complex than the gather-GEMM-scatter approach.

\subsubsection{Implicit GEMM Dataflow}
\label{sect:background:implicit_gemm}

\begin{figure}[!t]
    \centering
    \includegraphics[width=0.8\linewidth]{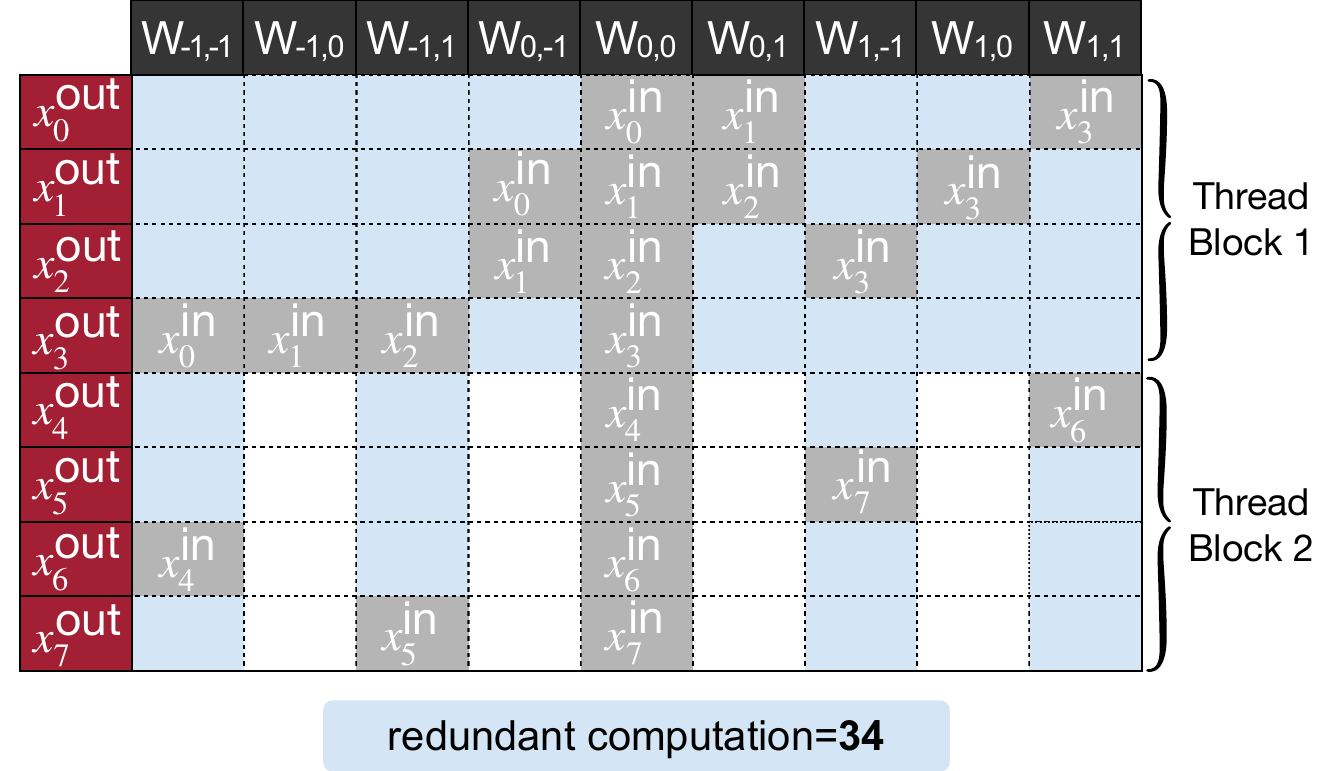}
    \caption{Illustration of the unsorted implicit GEMM dataflow for \fig{fig:background:workload} workload: each gray grid corresponds to a $C_\text{in}$-dimensional input feature and blue grids correspond to redundant computation. The input feature matrix is \textbf{not} stored in DRAM. We assume that each thread block contains 4 threads (4 rows).}
    \label{fig:background:implicit_gemm}
\end{figure}

\begin{figure}[!t]
    \centering
    \begin{subfigure}{\linewidth}
    \centering
    \includegraphics[width=0.8\linewidth]{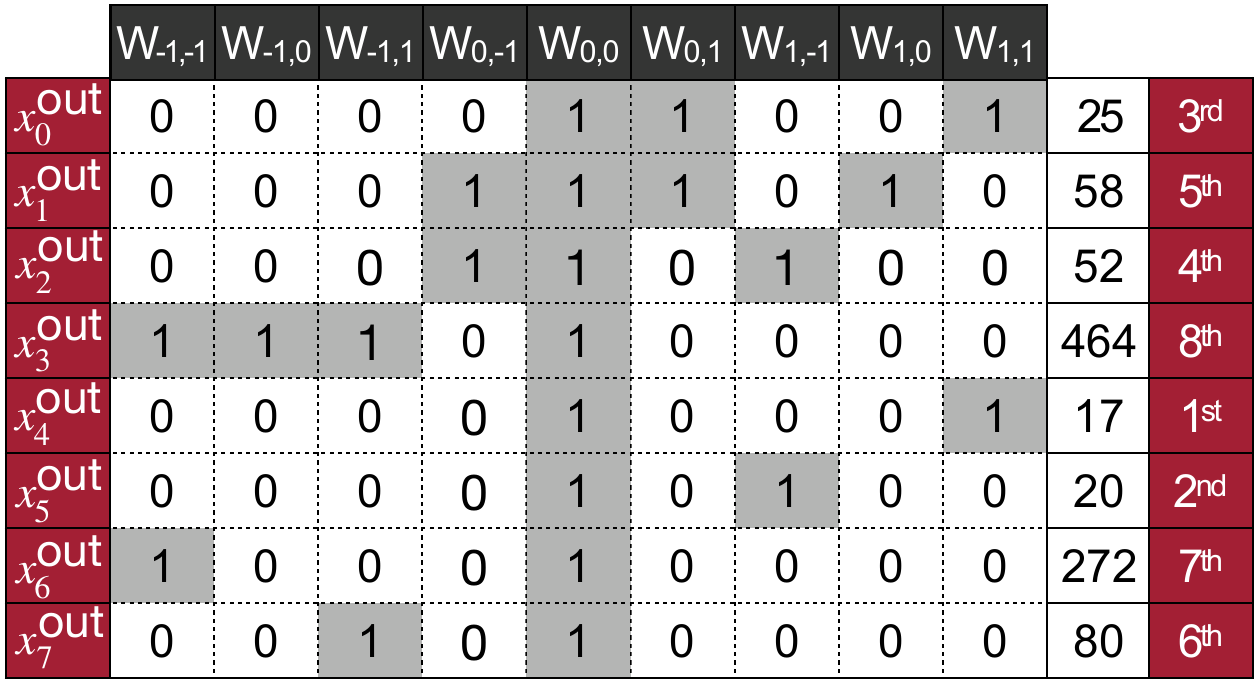}
    \caption{Bitmask in \SpConv v2.}
    \label{fig:background:sorted_implicit_gemm:bitmask}
    \end{subfigure}
    \hfill
    \begin{subfigure}{\linewidth}
    \centering
    \includegraphics[width=0.8\linewidth]{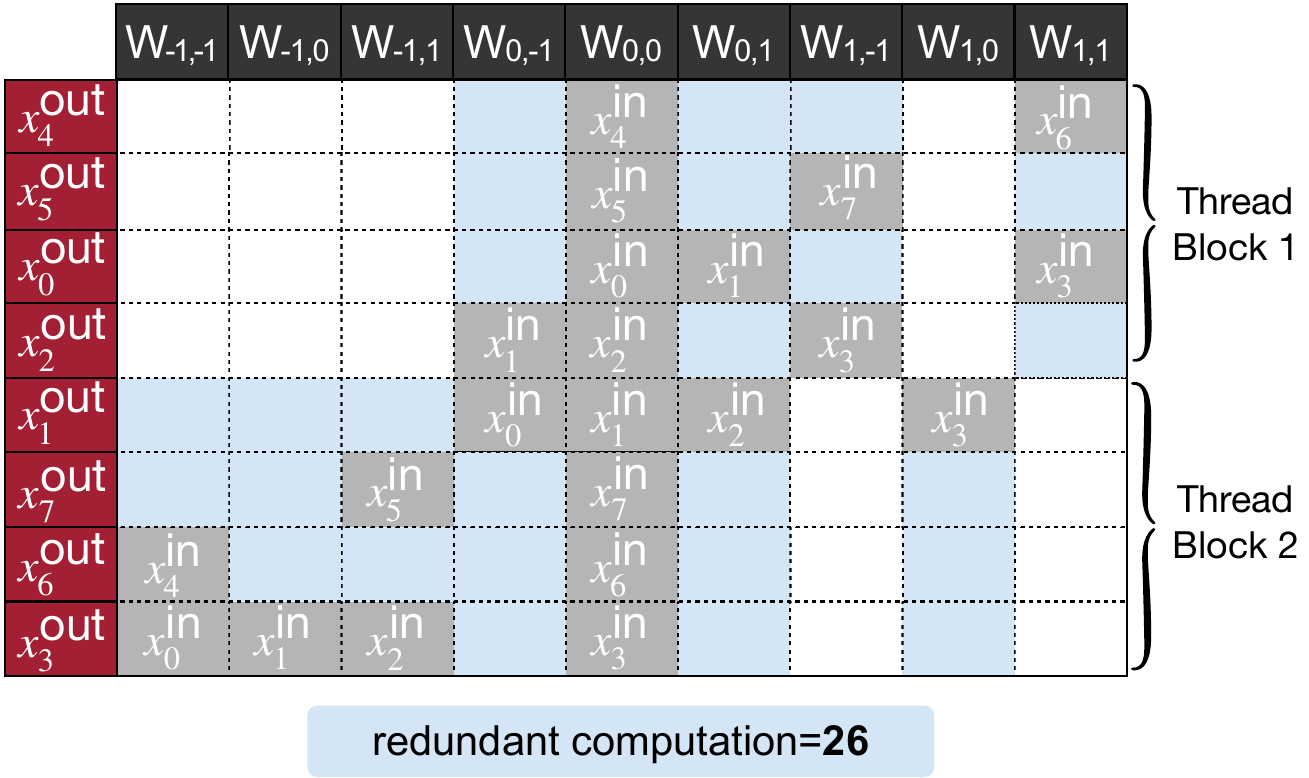}
    \caption{Sorting reduces redundant computation in \SpConv v2.}
    \label{fig:background:sorted_implicit_gemm:overhead}
    \end{subfigure}
    \caption{\SpConv v2 sorts the input bitmasks and reorders the computation accordingly. White grids are skipped zero computation. Consequently, redundant computation is reduced from 34 MACs (\fig{fig:background:implicit_gemm}) to 26 for the \fig{fig:background:workload} example.}
    \label{fig:background:sorted_implicit_gemm}
\end{figure}

Very recently, \SpConv v2~\cite{yan2018second,yan2022spconv} extends the well-known implicit GEMM formulation~\cite{chetlur2014cudnn,nvidia2022cutlass} for 2D convolution to 3D sparse convolution. As in \fig{fig:background:implicit_gemm}, the sparse convolution workload in \fig{fig:background:workload} is equivalent to a dense GEMM $\bm{X}^{\text{out}}=\bm{X}^{\text{im2col-in}}_{M\times K}\times \bm{W}_{K\times N}$. Here, $M=N_\text{out}$ (number of output points), $N=C_\text{out}, K=\lvert\Delta^D(K)\rvert C_\text{in}$. We visualize matrix $\bm{X}^{\text{im2col-in}}$ 
 in \fig{fig:background:implicit_gemm}, where we have $N_\text{out}\times \lvert\Delta^D(K)\rvert$ grids and each corresponds to $C_\text{in}$ channels. Gray grids corresponds to input features. For example, output point $\bm{q}_1$ has four neighbors: $\bm{p}_0$ (with $\bm{w}_{0,-1}$, $\bm{p}_1$ (with $\bm{w}_{0,0}$), $\bm{p}_2$ (with $\bm{w}_{0,1}$) and $\bm{p}_3$ (with $\bm{w}_{1,0}$) so the second row in \fig{fig:background:implicit_gemm} has four gray entries. One can verify that $\bm{x}^{\text{out}}_1 = \bm{X}^{\text{im2col-in}}_1\times\bm{W}$ against \fig{fig:background:workload}. 

Similar to fetch-on-demand, implicit GEMM overlaps computation with memory access (\fig{fig:background:waterfall}). This allows us to hide the memory latency through pipelining. Like \texttt{im2col} in 2D convolution, an implicit GEMM implementation is output-stationary. So it achieves the theoretical minimum DRAM write-back traffic. However, despite having lower DRAM traffic compared to fetch-on-demand, implicit GEMM has non-negligible redundant computation. As shown in \fig{fig:background:implicit_gemm}, we assume that each warp contains four threads. All GPU threads within a warp execute in lockstep. Whenever a thread has a non-empty neighbor at weight $\bm{\delta}$, all threads in the warp will either perform computation or waste cycles for that weight. This leads to 34 redundant computation in \fig{fig:background:implicit_gemm}, which is even more than 22 effective MACs in this example. 

To address this issue, SpConv v2 excludes unsorted implicit GEMM in their design space and utilizes bitmask sorting to minimize computation overhead. Following the approach taken by DSTC~\cite{wang2021dual}, each output point is assigned a $K^D$-dimensional bitmask that indicates the presence of its neighbors. These bitmasks are treated as numbers and sorted, and the order of computation for different outputs is adjusted accordingly. For instance, warp 0 calculates $\bm{x}_{0-4}^{\text{out}}$ in \fig{fig:background:implicit_gemm}, but it calculates $\bm{x}_{4,5,0,2}^{\text{out}}$ in \fig{fig:background:sorted_implicit_gemm}\textcolor{linkcolor}{b} instead.  Thanks to sorting, computation overhead is reduced from 34 MACs to 26 MACs. In practical applications, sorting can reduce redundant computation by up to 3$\times$, but it remains unclear whether this reduction translates into proportional speedups.

\subsection{Motivation}

As mentioned above, gather-GEMM-scatter is easy to implement but has poor performance. The more performant dataflows with overlapped computation and memory access cannot be implemented with the help of vendor libraries. Implementing the state-of-the-art implicit GEMM dataflow alone is a daunting task, as demonstrated by the SpConv v2 authors who had to painstakingly re-implement the entire CUTLASS framework from scratch with a custom Python-based template metaprogrammer~\cite{yan2022cumm}. The resulting code base has over \textbf{40,000} lines of code which increases the risk of errors for developers. This also makes it challenging for the community to explore a wider design space for sparse point cloud convolution kernels, hindering further performance improvements.

Therefore, in \system, we want to first demonstrate in \sect{sect:kernel_generator} that highly efficient dataflows with overlapped computation and memory access can be generated with a relatively low engineering complexity (comparable to implementing gather-GEMM-scatter). With the efficient kernel generator as a cornerstone, we further showcase in \sect{sect:autotuner} that the design space for sparse point cloud convolution could be significantly extended, and there exists solutions that are up to \textbf{1.7$\times$} faster in inference, \textbf{1.3$\times$} faster in training compared with the incumbent state-of-the-art within this vast space. Tackling a fundamentally \textit{sparse} workload, we also challenge traditional thinking on \textit{dense} GPU kernel design. Our research reveals that typical first-order performance indicators, such as total computation, DRAM access, or even total runtime for all sparse convolution \textbf{computation kernels}, cannot accurately reflect the \textbf{end-to-end runtime} of sparse point cloud workloads. This is because sparse workloads require expensive mapping operations. On top of this observation, we will further demonstrate that end-to-end optimal dataflows could sometimes choose configurations with up to \textbf{6$\times$} computation overhead and \textbf{4$\times$} larger DRAM footprint. %

\section{Sparse Kernel Generator}
\label{sect:kernel_generator}

\begin{table}[t]
\caption{Different sparse convolution dataflows in \sect{sect:background} can be mapped onto GPUs as dense GEMM with sparse global memory iterators.}
\scalebox{0.8}{
\begin{tabular}{lccc} \toprule
                    & \textbf{DRAM $\rightarrow$ L1 SRAM} & \textbf{MMA}   & \textbf{RF $\rightarrow$ DRAM} \\\midrule
GEMM                & dense                        & dense & dense                 \\
\textcolor{gray}{gather-GEMM-scatter} & \textcolor{gray}{dense + gather}                        & \textcolor{gray}{dense} & \textcolor{gray}{dense + scatter}                \\
fetch-on-demand     & \textbf{sparse}                       & dense & \textbf{sparse}                \\
implicit GEMM       & \textbf{sparse}                       & dense & dense \\\bottomrule               
\end{tabular}
}
\label{tab:method:summary}
\end{table}

\begin{figure}[!t]
    \centering
    \includegraphics[width=0.9\linewidth]{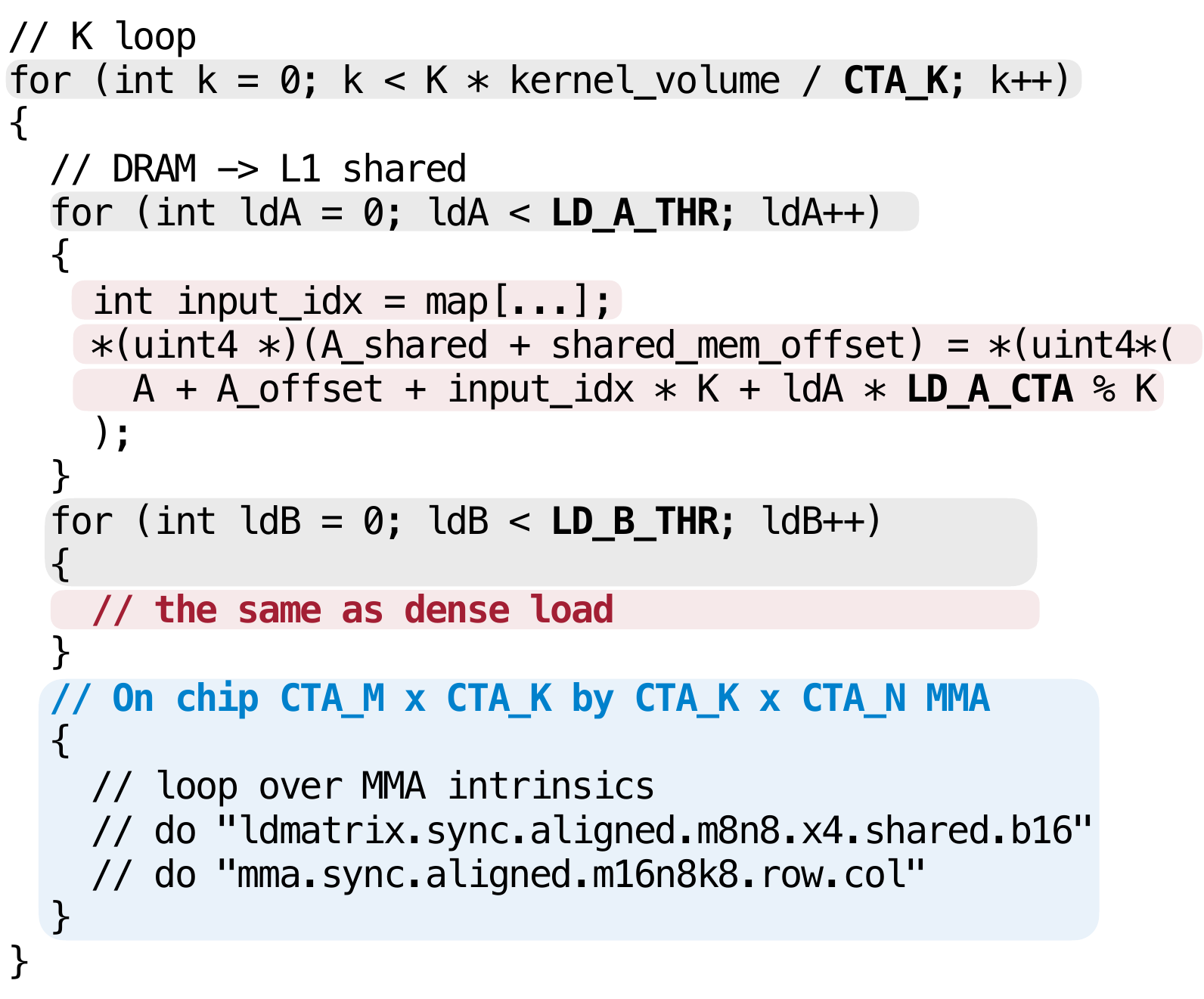}
    \caption{We introduce Sparse Kernel Generator, a code generator that integrates on-chip MMA subroutines from~\cite{chen2018tvm} directly at the source code level, unlocking the potential of using \textit{dense}, \textit{fixed shape} tensor compiler to generate programs for \textit{sparse}, \textit{dynamic shape} workloads. \textbf{\textcolor{gray}{Gray}}: constant code, \textbf{\textcolor{mydarkred}{red}}: fixed metaprogramming template, \textbf{\textcolor{mydarkblue}{blue}}: generated automatically by existing tensor compiler for each tile size.}
    \label{fig:method:sparse_kernel_generator_code}
\end{figure}

\begin{figure}
    \centering
    \includegraphics[width=\linewidth]{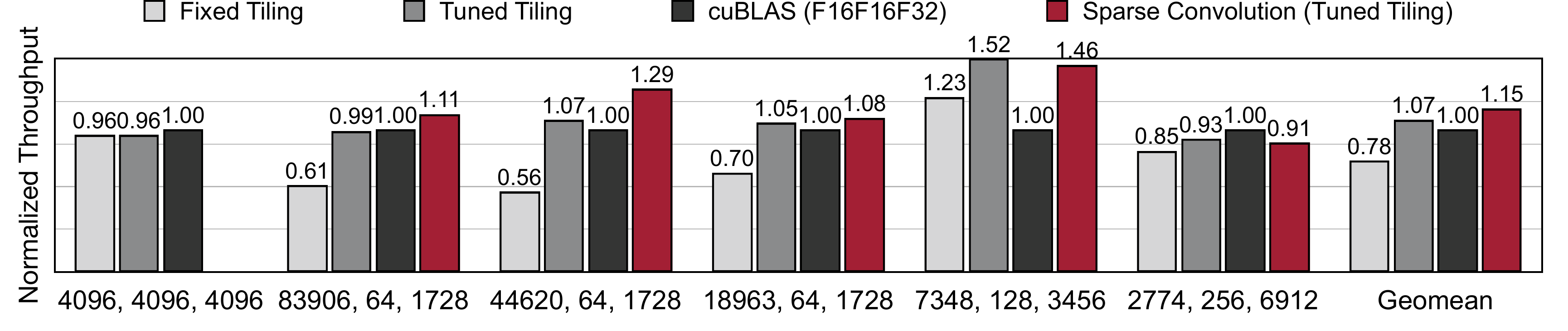}
    \caption{For sparse convolution workloads (MinkUNet on SemanticKITTI), it is possible for our template to achieve or even exceed cuBLAS utilization for the equivalent-sized GEMM problem by tuning \textit{only tiling size parameters}.}
    \label{fig:method:fixed_tiling}
\end{figure}

In this section, we introduce the Sparse Kernel Generator, which is a metaprogrammer that can efficiently generate sparse convolution GPU kernels. Existing metaprogrammers, such as TVM~\cite{chen2018tvm}, are designed to generate optimized GPU computing schedules for \textit{dense} and \textit{fixed-shape} workloads. However, point cloud workloads are naturally \textit{sparse} and have \textit{dynamic shapes}. %

\subsection{Dense to Sparse Adaptation}
\label{sect:method:dense_to_sparse}

Leveraging the information from \sect{sect:background}, we establish the relationship between sparse convolution and dense GEMM kernels, as summarized in \tab{tab:method:summary}. We show that the fetch-on-demand and implicit GEMM dataflows with their overlapped memory access and computation can be seen as generalized GEMM kernels with sparse DRAM loading and write-back iterators. Take implicit GEMM as an example, we start from its equivalent-sized dense GEMM workload in \sect{sect:background:implicit_gemm}. We notice that position $(i, j)$ in $\bm{X}^{\text{im2col-in}}$ is mapped to position $(\mathcal{M}_{i, j / C_\text{in}}, j\%C_\text{in})$ in $\bm{X}^{\text{in}}$. Here $\mathcal{M}_{N_\text{out}\times \lvert\Delta^D(K)\rvert}$ is the output-stationary representation of the maps defined in \sect{sect:background:gather_gemm_scatter}. For the $n$\textsuperscript{th} output point, if its $k$\textsuperscript{th} neighbor is non-empty, then $\mathcal{M}_{n,k}$ is the index of this neighbor; otherwise $\mathcal{M}_{n,k}=-1$. For example, in \mbox{\fig{fig:background:implicit_gemm}}, $\mathcal{M}_{2,3}=1$ since the fourth neighbor of $\bm{x}_2^{\text{out}}$ is $\bm{x}_1^{\text{in}}$. Here we assume indices start from 0. By introducing this one level of indirect addressing, we can easily transition from a dense GEMM to a sparse implicit GEMM when loading data from DRAM to L1 shared SRAM. Since the DRAM$\rightarrow$L1 memory access to $\bm{W}$ is dense, one can reuse the CUDA code segment for 2\textsuperscript{nd} operand loading in dense GEMM. Based on this formulation, as in \fig{fig:method:sparse_kernel_generator_code}, a sparse convolution kernel can then be decomposed into three parts. The \textbf{\textcolor{gray}{gray}} code is \textit{always constant}. \textbf{\textcolor{mydarkblue}{Blue}} code depends on the tile sizes and can be automatically generated by the existing compilers~\cite{chen2018tvm}. The \textbf{\textcolor{mydarkred}{red}} code cannot be generated by existing dense tensor compilers due to sparsity, but it can be generated from a \textit{fixed} template that only takes in tiling sizes as input parameters. Consequently, we only need to manually implement the short \textbf{\textcolor{mydarkred}{red}} code template and a TensorIR~\cite{feng2022tensorir} template that outputs the \textbf{\textcolor{mydarkblue}{blue}} on-chip MMA subroutine, which only takes hundreds of lines of code (orders of magnitude cheaper than the \SpConv v2 code generator). 

For simplicity, we did not visualize performance optimization techniques such as double buffering and pipeling in \fig{fig:method:sparse_kernel_generator_code}. However, these techniques will not impact the design of our code generator. Similar analysis and code transformation can also be applied to the fetch-on-demand dataflow. %

\subsection{Static to Dynamic Adaptation}
\label{sect:kernel_generator:static_to_dynamic}

Thanks to the adaptation described in \sect{sect:method:dense_to_sparse}, we can now easily implement sparse convolutions in dataflows with overlapped computation and memory access. However, the simplicity of the code generator comes at the cost of a reduced design space. Our Sparse Kernel Generator only allows the \textit{tiling sizes} to be tuned, while leaving most of the dimensions in the tensor program design space to be fixed (\eg the order and split of the loop nests). Fortunately, we argue that such reduced design space does not compromise the performance. We present an idealized experiment in \mbox{\fig{fig:method:fixed_tiling}}. We manually traverse all possible tile sizes for different layers in MinkUNet~\cite{choy20194d} on SemanticKITTI~\cite{behley2019semantickitti} and apply compile-time constant folding to maximize performance. We benchmark the resulting sparse kernel with the lowest latency against cuBLAS, which runs an equivalent-sized GEMM problem due to the lack of sparsity support. It turns out that we can achieve $>100\%$ cuBLAS utilization on average by only tuning tile sizes. Notably, for the last workload, the equivalent-sized dense GEMM problem can run at $\approx$90\% device utilization on RTX 3090. If we ignore redundant computation (\fig{fig:background:implicit_gemm}), it is safe to assert that extending the design space beyond tile sizes will not significantly improve final performance on this workload. 

Despite achieving encouraging results in the idealized experiment, it remains challenging to transfer the performance to real systems. Unlike dense workloads, each sparse point cloud sample has a different shape in terms of the number of points. Pre-compiling constant-folded kernels for all possible workloads, as is done by TVM and TensorRT in the dense domain, is impossible for us. Naively unfold the constants in fixed shape kernels and revert them back to workload shape parameters will degrade the performance by up to \textbf{1.7$\times$}. This totally undermines the good results achieved in \fig{fig:method:fixed_tiling}. Worse still, the first \textbf{\textcolor{mydarkred}{red}} instruction in \fig{fig:method:sparse_kernel_generator_code} now requires explicit boundary check in flexible shape kernels, which brings up to \textbf{1.35$\times$} performance overhead as well.

To this end, we present two simple yet effective strategies to address these two performance roadblocks. 

We first pinpoint that the slow addressing of $\bm{X}^\text{in}$ is the reason why constant unfolding ruins the performance. Unlike in dense GEMM, accessing $\bm{X}^\text{in}$ requires two inefficient division and modulo operations with $C_\text{in}$ as an operand, which are necessary just for addressing. This impacts the efficiency since $C_\text{in}$ is stored in the RF and has an access latency no shorter than L1 on GPUs. Worse still, accesses to $\bm{X}^\text{in}$ are located in the innermost layer of the long $K$ loop ($\lvert\Delta^D(K)\rvert\times C_\text{in}$, ranging from 1728 to 6912 in \fig{fig:method:fixed_tiling}). Fortunately, we notice that most of the addressing computation is irrelevant to the innermost loop variable \texttt{ldA} in \fig{fig:method:sparse_kernel_generator_code}. Therefore, it is possible for us to lift the loop invariants out of the loop. For real tiling sizes with \texttt{LD\_A\_THR}=4 and 8, this at least reduces addressing cost by 4-8$\times$. We further analyze the template and perform loop invariant hoisting wherever possible. Ablation studies in \sect{sect:analysis:sparse_kernel_generator} shows that addressing simplification can fully close the up to 1.7$\times$ constant unfolding overhead.

Likewise, among all boundary checks in the dynamic shape kernel, the one for accessing \texttt{map} within the innermost \texttt{ldA} loop is the most time-consuming. Although loop invariant hoisting does not apply in this case, we can solve this issue by padding the first dimension of \texttt{map} to be a multiple of \texttt{cta\_M}. With this simple modification, no boundary check on \texttt{map} access in \fig{fig:method:sparse_kernel_generator_code} is required since we can ensure that every access stays within bounds. With that reduced control flow overhead, we close the final 1.14-1.35$\times$ performance gap between fixed and dynamic shape kernels.

\section{Sparse Autotuner}
\label{sect:autotuner}
\begin{figure}
    \centering
    \includegraphics[width=\linewidth]{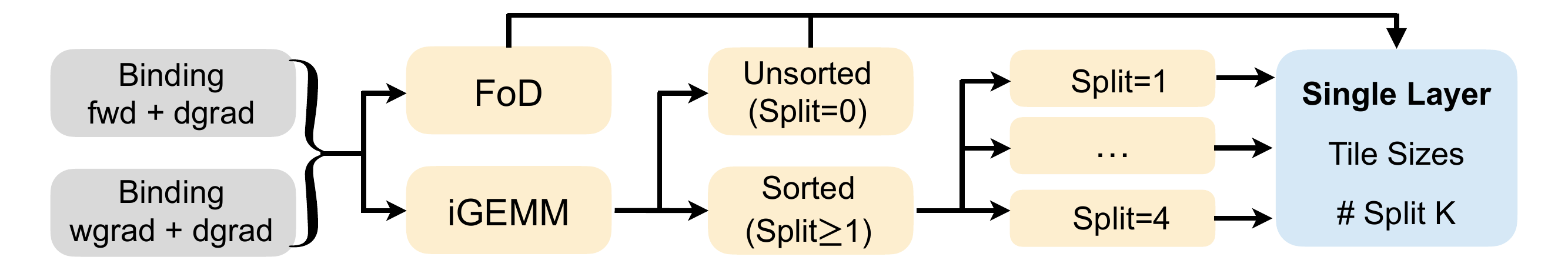}
    
   \caption{Overview of \system design space.}
    \label{fig:method:design_space}
\end{figure}

Based on the simple yet powerful Sparse Kernel Generator, we present Sparse Autotuner. It first significantly enlarges the design space of existing libraries (illustrated in \fig{fig:method:design_space}) and then applies group-based configuration tuning across this enlarged space.

\subsection{Design Space Augmentation}

\begin{figure}[!t]
    \centering
    \includegraphics[width=0.8\linewidth]{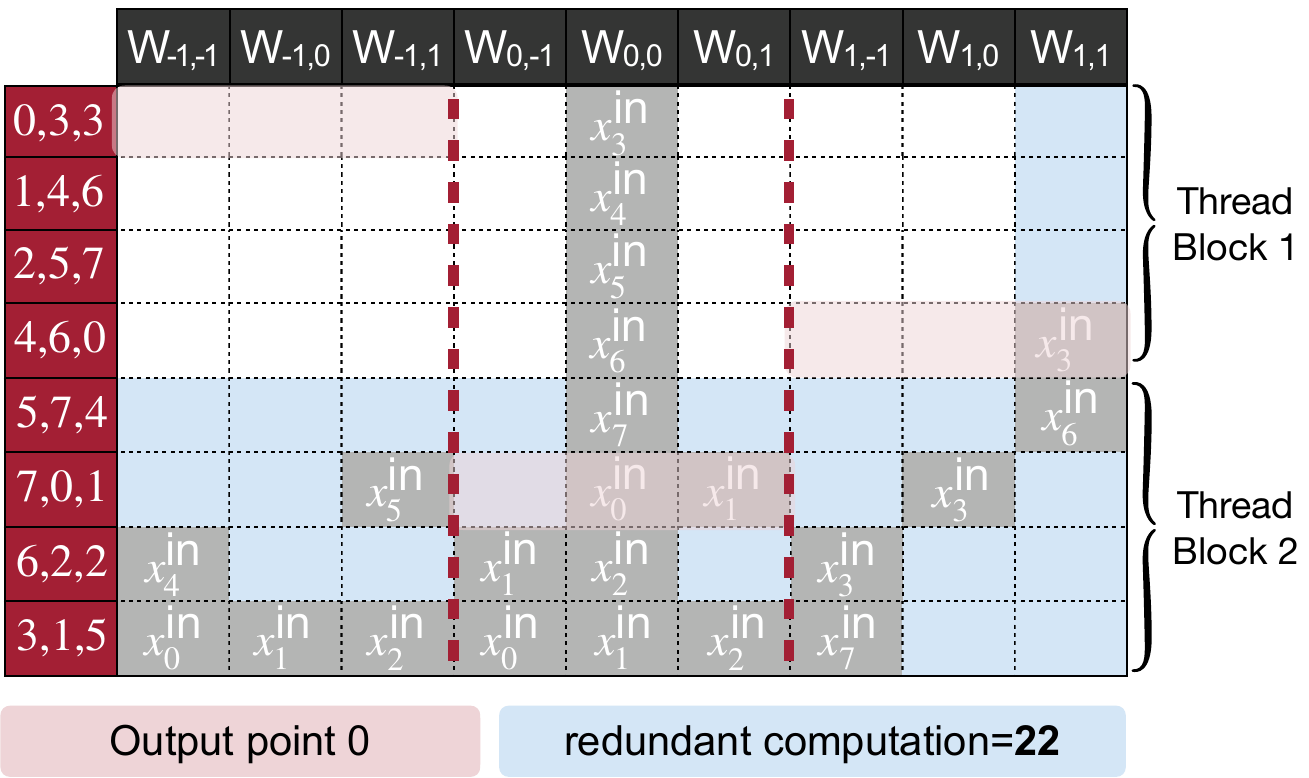}
    \caption{We extend the implicit GEMM design space by introducing arbitrary number of mask splits. Compared with \fig{fig:background:sorted_implicit_gemm:overhead} (1 split), splitting the mask into three parts further reduces redundant computation and increases parallelism.}
    \label{fig:method:more_split_sort_igemm}
\end{figure}

\begin{figure}
    \centering
    \includegraphics[width=\linewidth]{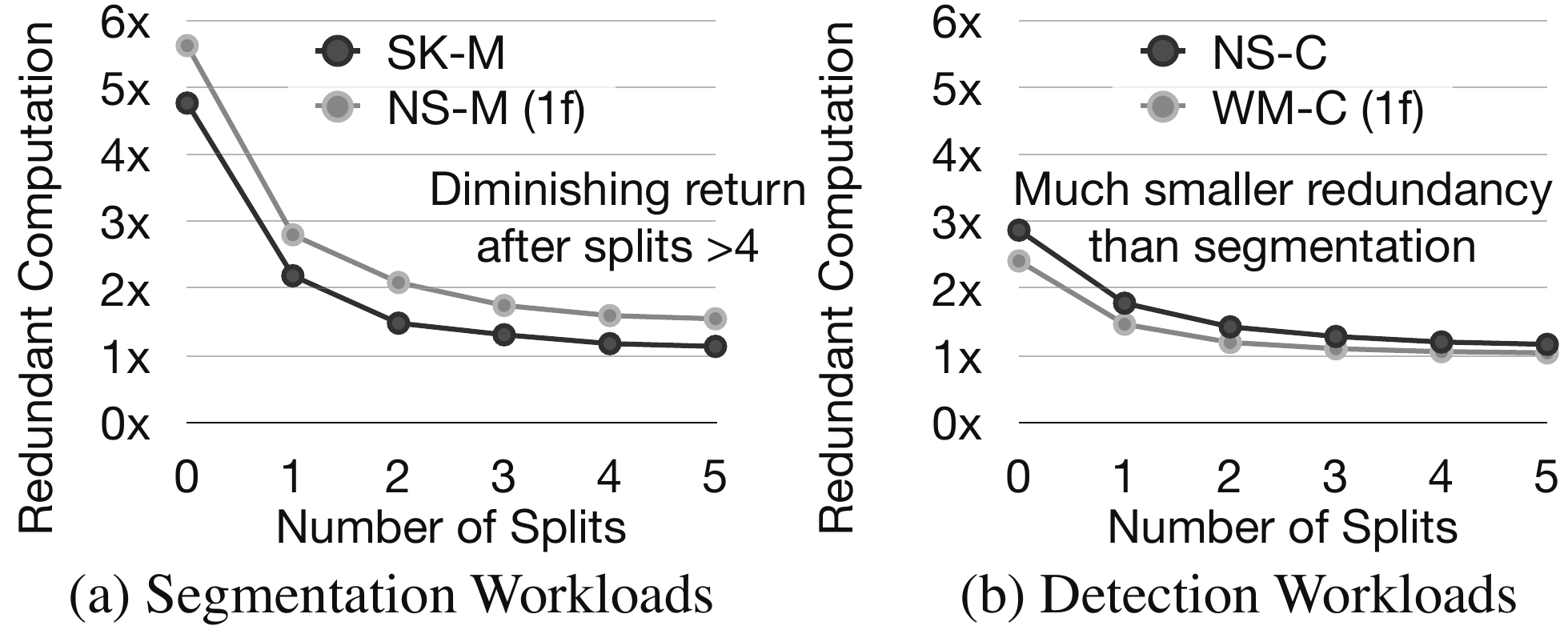}
    \caption{A large design space on number of splits in implicit GEMM is beneficial: (a) redundant computation in segmentation workloads continues to drop quickly until splits = 5; (b) redundant computation in detection workloads at split = 0 (unsorted) is acceptable on high-parallelism devices.}
    \label{fig:method:num_of_splits}
\end{figure}

Thanks to the simplicity of Sparse Kernel Generator, we can easily expand our design space. Since the generator can produce fetch-on-demand kernels, we can effortlessly incorporate this dataflow in our designs. Besides, for implicit GEMM, number of splits (\fig{fig:method:more_split_sort_igemm}) is an important tunable dimension in the implicit GEMM dataflow that was previously overlooked. Similar to the SplitK technique~\cite{nvidia2022cutlass} in dense GEMM kernel design, one could split the sequential $K$ loop in \fig{fig:method:sparse_kernel_generator_code} into $s$ parts. By doing so, each split (whose $K$ loop is now $s\times$ shorter) can compute in parallel and write to a separate DRAM buffer. These partial sums are later reduced by a summation kernel to produce the final result. We also reorder the computation in each split following \fig{fig:background:sorted_implicit_gemm}, which involves argsorting $s$ individual bitmasks and reordering the \texttt{map} accordingly. For example, after reordering, the first row calculates part of $\bm{x}_0^{\text{out}}, \bm{x}_3^{\text{out}}, \bm{x}_3^{\text{out}}$, while the full feature of $\bm{x}_0^{\text{out}}$ is calculated in the 1\textsuperscript{st}, 4\textsuperscript{th}, 6\textsuperscript{th} rows by two thread blocks collaboratively. As such, there are more common zero neighbors for each thread block and the redundant computation is further reduced from 26 in \fig{fig:background:sorted_implicit_gemm} to 22 in \fig{fig:method:more_split_sort_igemm}. When integrating support for arbitrary split 
implicit GEMM, we notice that it is beneficial to reorder the \texttt{map} in an offline manner for a similar reason in \sect{sect:kernel_generator:static_to_dynamic}.

Conventionally, dense GPU kernel design is often guided by first-order performance approximation (\mbox{\eg} computation and DRAM footprint). Following these proxies, it seems to be reasonable to eliminate split = 0 (unsorted implicit GEMM in \mbox{\fig{fig:background:implicit_gemm}}) due to its large redundant computation. Split > 2 should also be eliminated since it incurs much larger DRAM write back traffic. In fact, such prematured optimizations lead to the restricted design space in \SpConv v2. However, we argue in \fig{fig:method:num_of_splits} that it is beneficial to have a larger design space that includes many first-order suboptimal solutions. On the one hand, the redundant computation in both segmentation and detection workloads keeps dropping until $s=5$. The difference in computation overhead between $s=2$ and $s=4$ can still be up to 1.2$\times$ for detection and 1.3$\times$ for segmentation. Thus, for devices with limited parallelism, it is beneficial to increase the number of splits despite increased DRAM traffic. On the other hand, when running detection workloads on devices with high parallelism, a 2.4-2.9$\times$ computation overhead for the unsorted dataflow in \fig{fig:background:implicit_gemm} is completely acceptable. We will demonstrate in \tab{tab:ablations:unsorted} and \tab{tab:ablations:unsorted_kernel} that kernels for detection will not run faster despite having $\sim2\times$ lower computation overhead on RTX 3090, which has an ample 71 TFLOPS FP16 peak throughput.

\subsection{Group-Based Configuration Tuning}

\begin{figure}[!t]
    \centering
    \includegraphics[width=0.9\linewidth]{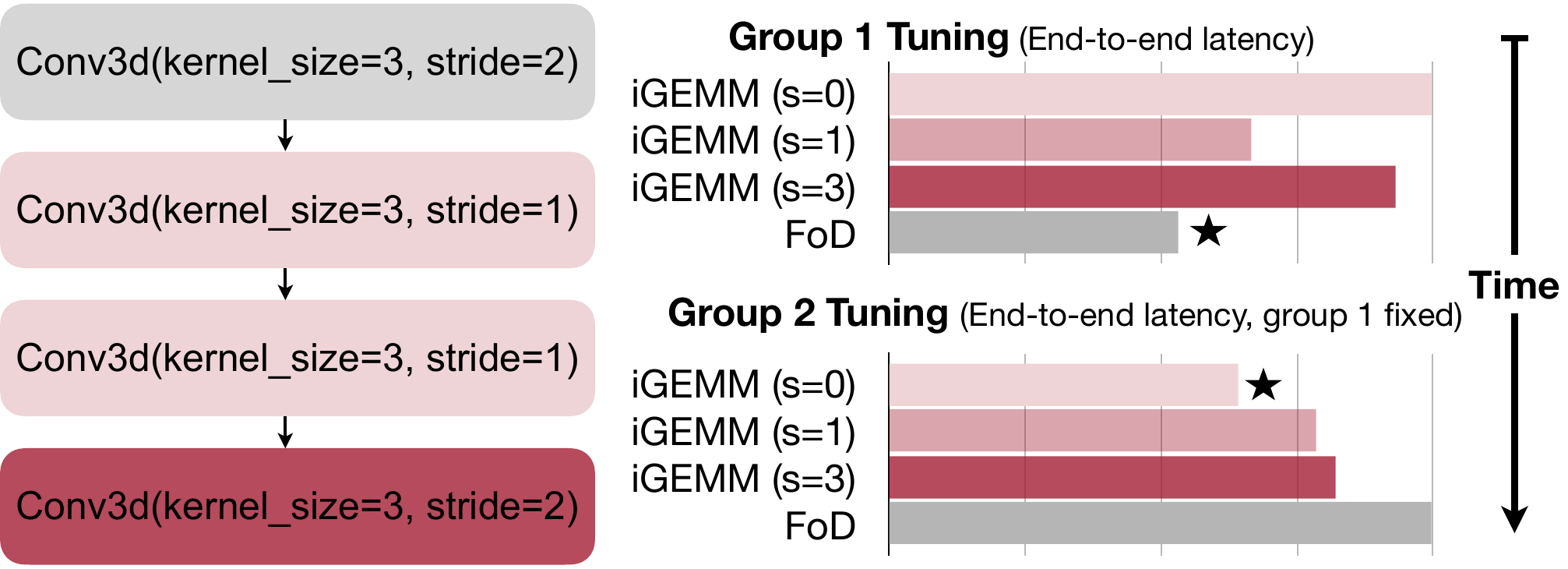}
    \caption{\textbf{Group-based autotuning}: Layers using the same maps will be assigned to the same group. After group partition, we exhaustively traverse all choices in our design space in a group-by-group manner and selects the best group configuration that leads to the lowest end-to-end latency.}
    \label{fig:background:autotuner}
\end{figure}

\begin{figure}[!t]
    \centering
    \includegraphics[width=\linewidth]{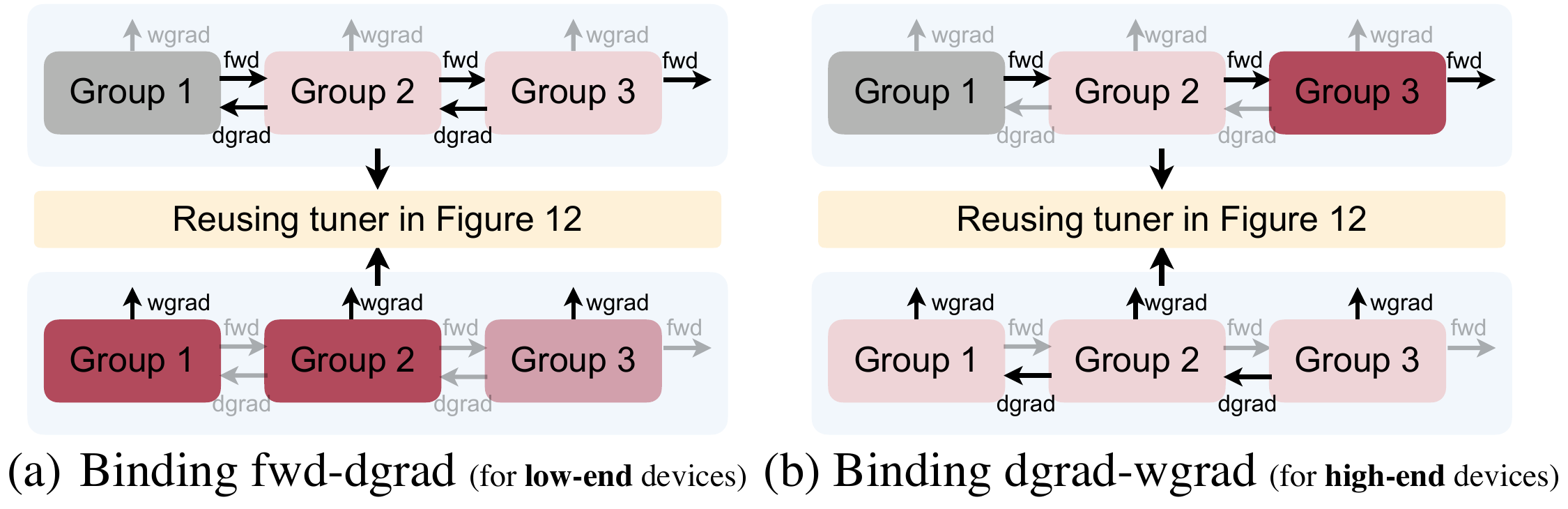}
    \caption{\textbf{Parameter binding in training tuner}: we propose to partially decouple the dataflow parameters for \texttt{forward}, \texttt{dgrad} and \texttt{wgrad} kernels in training, which leads to up to 10\% improvement in end-to-end training time.}
    \label{fig:background:autotuner_bind}
\end{figure}

To this end, we designed a \mbox{\textit{sparse}} and \mbox{\textit{dynamic shape}} kernel generator with minimal help from \mbox{\textit{dense}} and \mbox{\textit{fixed shape}} tensor compilers. By doing so, we obtain high-performance sparse convolution kernels with different dataflows (\eg fetch-on-demand and implicit GEMM) and augments the design space of implicit GEMM itself by introducing arbitrary number of mask splits. However, no dataflow is perfect for all workloads. As in \sect{sect:background}, fetch-on-demand has zero redundant computation but suffers from large DRAM scattering traffic, while implicit GEMM has the exact opposite property. Similarly, there is no single set of parameters that works for each dataflow. For example, the number of splits $s$ in implicit GEMM reflects the tradeoff between redundant computation and control flow overhead (\eg sorting $s$ individual bitmasks and reordering the maps). Therefore, the enlarged design space necessitates the design of an autotuning system that can automatically determine the optimal dataflow and dataflow-specific parameters for different workloads.

To determine the optimal dataflow for different layers, we divide all layers into different groups (illustrated in \fig{fig:background:autotuner}). All layers within each group use the same input-output mappings (\textit{maps}) and are forced to execute the same dataflow. This is because different dataflows require different map structures. Implementations such as gather-GEMM-scatter and fetch-on-demand require the maps to be stored in a weight-stationary order, represented as $\mathcal{M}_{\bm{\delta}} = \{(\bm{p}_j, \bm{q}_k)\lvert\bm{p}_j=s\bm{q}_k+\bm{\delta},\bm{q}_k\in\bm{P}^{\text{out}}\}$, which makes it difficult to infer all the neighbors of an output point (required by implicit GEMM). On the other hand, the implicit GEMM implementation, stores the maps in an output-stationary order, represented as $\mathcal{M}_k = \{\bm{p}_k^{(\bm{\delta})}\lvert \bm{p}_k^{(\bm{\delta})}=s\bm{q}_k+\bm{\delta},\bm{\delta}\in\Delta^D(K)\}$, which makes it difficult to infer all the inputs that use the same weight (required by the other two dataflows). It would incur significant overhead ($\sim$ latency of up to 3-4 sparse convolution layers within each group!) if we generate maps for all dataflows but only use one of them at runtime. %
Therefore, allowing intra-group heterogeneous dataflow selection is not desired. 
After group partition, we apply a group-level exhaustive search on a random subset of the target workload (\eg 100 scenes on the Waymo dataset). Since the execution time of each group is independent of the others, we tune the dataflow parameters in a greedy manner. We iterate over all possible choices for the $k$\textsuperscript{th} group based on the optimally-tuned configurations for the 1\textsuperscript{st} to $(k-1)$\textsuperscript{th} groups, using default parameters for all subsequent groups. This approach effectively reduces the tuner complexity from exponential to linear\footnote{Measuring end-to-end latency during tuning is necessary because in U-net structured models, some groups may contain layers that are not consecutive in the original model.} and allows us to complete tuning within 2 minutes for most workloads. Considering that the tuned schedule could be reused for millions of scenes in real-world ADAS applications during inference, the cost is clearly justifiable.

We further extend Sparse Autotuner to support training workloads. The most straightforward design assumes that the back propagation kernels (\ie \texttt{dgrad} for feature map gradient calculation and \texttt{wgrad} for weight gradient calculation) share the same dataflow parameters as the \texttt{forward} kernel. However, as analyzed in \sect{sect:analysis:design_space}, such design incurs up to 10\% performance regression in end-to-end training. Naively decoupling the tuning process for training workloads leads to an unacceptable $O(K^3)$ tuning complexity, with $K$ being the size of our design space. To address this complexity issue, we partially bind dataflow parameters for \texttt{forward}, \texttt{dgrad}, and \texttt{wgrad} kernels. We propose two binding schemes: the \textbf{workload-pattern oriented} scheme binds the dataflow parameters for \texttt{forward} and \texttt{dgrad} kernels while allowing \texttt{wgrad} kernels to be tuned separately, reducing the tuning complexity to $O(K^2)$ and minimizing the total latency for all sparse convolution kernels. We also propose the \textbf{sparse-mapping oriented} scheme, which binds \texttt{dgrad} and \texttt{wgrad} kernels together since they share the same maps, minimizing the overhead for map computation. Similar to our observations in inference kernel autotuning, the high-parallelism devices (\eg A100) is far less sensitive to redundant computation than to mapping overhead, while the low-parallelism devices (\eg 2080 Ti) behaves in the exact opposite way. This explains our design choice to use scheme 1 for low-end devices and scheme 2 for more powerful GPUs. As a final remark, we further notice in \fig{fig:background:autotuner_bind} that the tuning time could be further reduced from $O(K^2)$ to $O(K)$ if we reuse the group-based tuner in \fig{fig:background:autotuner} twice and skip different parts of the kernels with dummy initializations during tuning.

\section{Evaluation}
\label{sect:evaluation}
\begin{figure*}[t]
    \centering
    \includegraphics[width=\linewidth]{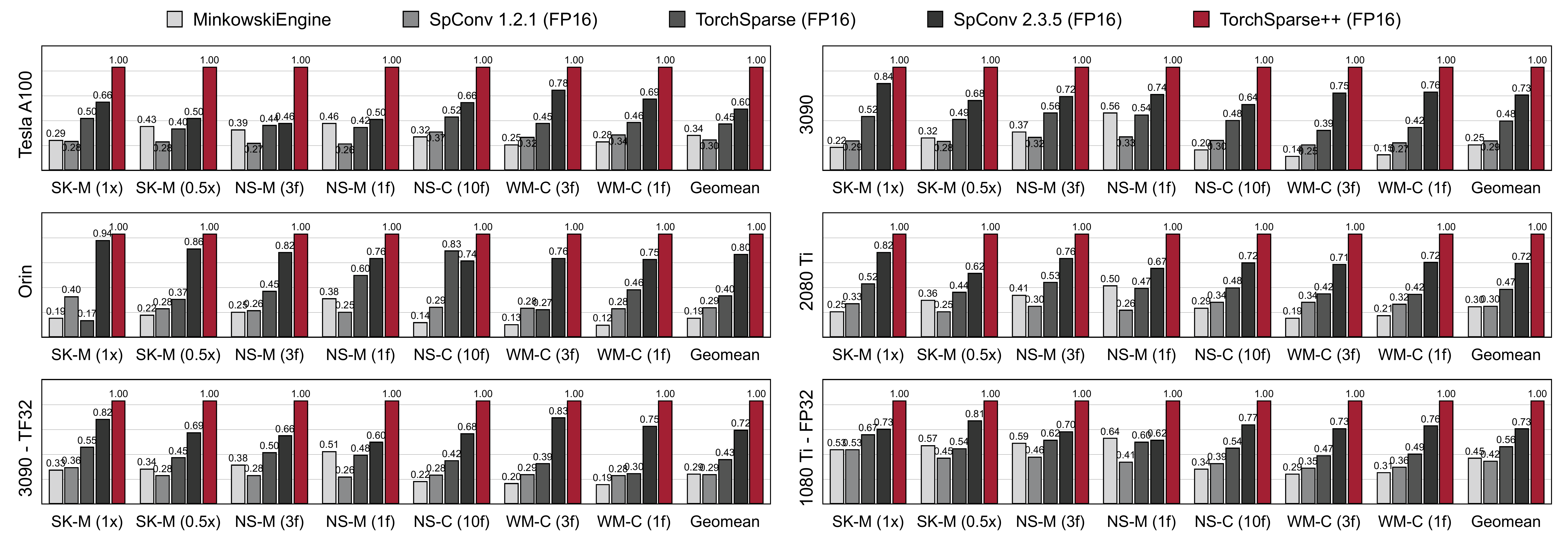}   
    \caption{\textbf{Inference Speedup:} \system significantly outperforms existing point cloud inference engines in both 3D object detection and LiDAR segmentation benchmarks across three generations of GPU architecture (Pascal, Turing and Ampere) and three precisions (FP16, TF32, FP32). It is up to \textbf{1.7}$\times$ faster than state-of-the-art \SpConv 2.3.5 and is up to \textbf{2.2$\times$} faster than \systemv1 on cloud GPUs. It also improves the latency of \SpConv 2.3.5 by \textbf{1.25$\times$} on Orin.} 
    \label{fig:evaluation}
\end{figure*}

\begin{figure*}[t]
    \centering
    \includegraphics[width=\linewidth]{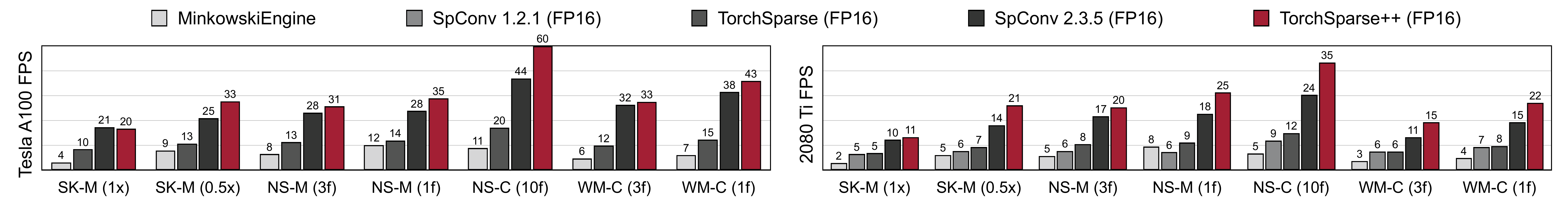}    
    \caption{\textbf{Training Speedup:} \system achieves faster FP16 training speed compared with \ME, \systemv1 and \SpConv 2.3.5. It is \textbf{1.16$\times$} faster on Tesla A100, \textbf{1.27$\times$} faster on RTX 2080 Ti than state-of-the-art \SpConv 2.3.5. It also significantly outperforms \ME by \textbf{4.6-4.8$\times$} across seven benchmarks on A100 and 2080 Ti.}
    \label{fig:evaluation:training}
\end{figure*}

\subsection{Setup}

We implement \system in CUDA and PTX assembly based on TorchSparse~\cite{tang2022torchsparse} and compare it with four state-of-the-art sparse convolution libraries \ME 0.5.4~\cite{choy20194d}, \SpConv 1.2.1~\cite{yan2018second}, \systemv1~\cite{tang2022torchsparse} (gather-GEMM-scatter) and \SpConv 2.3.5~\cite{yan2018second} (sorted implicit GEMM, released in April 2023).%
All systems are integrated into PyTorch 1.12.0 with CUDA 11.7 and cuDNN 8.4.1 and latency numbers are measured on NVIDIA GPUs.

We follow \systemv1~\cite{tang2022torchsparse} to evaluate all systems on seven representative real world 3D deep learning workloads: MinkUNet~\cite{choy20194d} (0.5$\times$/1$\times$ width, \texttt{SK-M}) on SemanticKITTI~\cite{behley2019semantickitti}, MinkUNet (1 or 3 frames, \texttt{NS-M}) on nuScenes-LiDARSeg~\cite{caesar2020nuscenes}, CenterPoint~\cite{yin2021center} (10 frames, \texttt{NS-C}) on nuScenes detection and CenterPoint (1 or 3 frames, \texttt{WM-C}) on Waymo Open Dataset~\cite{sun2020scalability}. Among these benchmarks, the SemanticKITTI and Waymo datasets are collected using 64-beam LiDAR sensors while the nuScenes dataset are collected using cheaper 32-beam LiDAR sensors. Multi-frame models increase LiDAR density by superimposing history LiDAR scans to the current frame. For detection workloads (CenterPoint), \textit{we only evaluate the runtime of \SparseConv layers}. 

\subsection{Results}

\paragraph{Inference.} We compare our results with the baseline designs including \ME, \SpConv 1.2.1, \systemv1 and \SpConv 2.3.5 in \fig{fig:evaluation}. All evaluations are done in unit batch size. \system consistently outperforms \textbf{all baseline systems} on GPUs with \textbf{all architectures} under \textbf{three numerical precisions} by a large margin. On cloud Ampere GPUs (A100 and 3090), it achieves \textbf{2.9-3.7}$\times$, \textbf{3.2-3.3}$\times$, \textbf{2.0-2.2}$\times$ and \textbf{1.4-1.7}$\times$ measured end-to-end speedup over the state-of-the-art MinkowskiEngine, SpConv 1.2.1, TorchSparse and SpConv 2.3.5, respectively. We also compare \system with \SpConv 2.3.5 on NVIDIA Jetson Orin, an edge GPU platform widely deployed on real-world autonomous vehicles. Our \system is \textbf{1.25$\times$} faster than \SpConv 2.3.5 on average, while achieving \textbf{1.3-1.4$\times$} consistent speedup across all detection workloads that are most time-critical in real ADAS applications. In addition, \system is competitive on legacy GPU architectures (Turing and Pascal), achieving \textbf{at least} \textbf{1.4$\times$}, \textbf{1.8$\times$}, \textbf{2.4$\times$}, \textbf{2.2$\times$} speedup over \SpConv 2.3.5, \systemv1, \SpConv 1.2.1 and \ME. Notably, recent advances in point cloud transformers~\cite{sun2022swformer,liu2023flatformer,wang2023dsvt} often claim superior accuracy-latency tradeoffs over sparse convolutional backbones implemented with the \SpConv v2 backend. With the much faster \system backend, assuming that the 2D part is deployed with TensorRT, the 3-frame CenterPoint model on Waymo is \textbf{1.5$\times$} faster than FlatFormer~\cite{liu2023flatformer} with higher accuracy on Orin.

\paragraph{Training.} We also compare the training performance of our \system and existing systems on A100 and 2080 GPUs in \fig{fig:evaluation:training}. We run the forward and backward pass of all workloads with batch sizes of 2 in mixed precision training (\ie all gradients are calculated in FP16 precision) except for \ME that does not support FP16. We make sure that all workloads evaluated in \fig{fig:evaluation:training} can reach the same accuracy using the \system backend compared with \systemv1 (for segmentation workloads) and \SpConv 2.3.5 (for detection workloads) with FP32 precision. Given the fact that A100 FP16 tensor core arithmetics has \textbf{16$\times$} higher throughput compared with FP32 (non-tensor core) computation (312 TFLOPS \vs 19.5 TFLOPS), we do not perform FP32 evaluation. As a result, \system is \textbf{4.6-4.8}$\times$, \textbf{2.5-2.6}$\times$ and \textbf{1.2-1.3}$\times$ faster than MinkowskiEngine, \systemv1 and SpConv 2.3.5 on both Ampere and Turing GPUs. \system paves the way for rapid model iteration for real-world ADAS applications. 

\begin{figure}[!t]
    \centering
    \includegraphics[width=\linewidth]{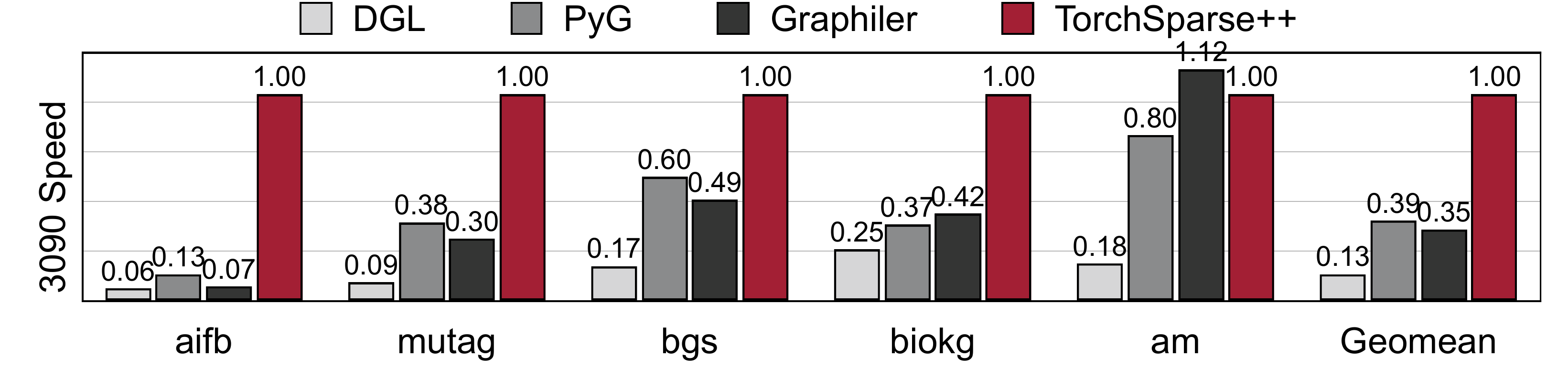}
    \caption{
    \mbox{\system} is \textbf{2.6-7.6$\times$} faster than DGL, PyG and Graphiler on five heterogeneous graph benchmarks.}
    \label{fig:experiments:evaluation_graph}
\end{figure}

\begin{table}[t]
\setlength{\tabcolsep}{3pt}
  \centering
  \renewcommand{\arraystretch}{0.85}
    \begin{tabular}{lccc}
    \toprule
    \textbf{Chip} & \textbf{RTX 3090} & \textbf{PointAcc} & \textbf{PointAcc-L} \\
    \midrule
    \textbf{System} & \system & ASIC & ASIC \\
    \textbf{Cores} & 328 & 64$^2$=4096 & 128$^2$=16384 \\
    \textbf{MACs} & 328$\times$64=20992 & 4096 & 16384\\
    \textbf{Frequency} & 1.7 GHz  & 1 GHz & 1 GHz \\
    \textbf{Peak Performance} & 35.5 TMACS & 4 TMACS & 16 TMACS\\\midrule
    \textbf{Proj. Latency} & 31.6 ms & -- & 17.8 ms\\
    \bottomrule
    \end{tabular}%
  \caption{We scale up the systolic array of PointAcc~\cite{lin2021pointacc} to match the peak performance of RTX 3090 and compare our \system against the ASIC accelerator.}
  \label{tab:asics}%
\end{table}%

\paragraph{Comparison against Accelerators.}
We further compare the performance of \system on RTX 3090 against a scaled-up version of PointAcc~\cite{lin2021pointacc} using the SemanticKITTI-MinkUNet workload. The systolic array in PointAcc is enlarged from 64$\times$64 to 128$\times$128 to roughly match the number of MACs (16384 \vs 20992) on RTX 3090. The PointAcc memory bandwidth is scaled up accordingly. Since the accelerator adopts \texttt{IC-OC} parallelism, we assume that the scaled PointAcc-L achieves linear speedup if the executed layer has large enough input and output channels. We also scale the measured \system latency by 1.7 (clock frequency difference) $\times$ 1.3 (peak MACs difference) = 2.2$\times$ for fair comparisons. As a result, \system achieves \textbf{56\% of ASIC speed} on a general-purpose hardware platform with similar computation budget. Notice that we also attempt to make a direct comparison with Mesorasi~\cite{feng2020mesorasi}, which codesigns the point cloud convolution algorithm with the hardware architecture. However, its delayed aggregation scheme could only work for convolution operators with \textit{shared} weights for all neighbors. The main workload accelerated in this paper, sparse convolution, is more complicated because it has different weights for different neighbors (see \fig{fig:background:workload}). Therefore, such comparison might be hard to achieve. 

\paragraph{Results on Graph Workloads}. We also implement R-GCN~\cite{schlichtkrull2018modeling} with TorchSparse++ and benchmark it on five representative heterogeneous graph datasets against state-of-the-art graph deep learning systems DGL~\cite{wang2019deep}, PyG~\cite{fey2019fast} and the Graphiler~\cite{xie2022graphiler} compiler. \system is \textbf{7.6$\times$}, \textbf{2.6$\times$}, \textbf{2.9$\times$} faster and \textbf{3.4$\times$}, \textbf{4.4$\times$}, \textbf{5.6$\times$} more memory efficient compared with DGL, PyG and Graphiler.

\section{Analysis}
In this section, we present in-depth analysis on the design choices of our Sparse AutoTuner and Sparse Kernel Generator and ablate the source of performance gains in \sect{sect:evaluation}.

\subsection{Design Space of Sparse AutoTuner}
\label{sect:analysis:design_space}

\begin{table}
\centering\small

\begin{tabular}{lccc} \\\toprule
\textbf{nuScenes-CenterPoint} & Unsorted & Split=1 & Split=2 \\ \midrule
RTX 3090 latency (ms)                     & 7.45     & 8.18    & 8.81    \\
Orin latency (ms)                         & 21.90    & 22.20   & 23.93   \\\midrule
\textbf{Waymo-CenterPoint-1f} & Unsorted & Split=1 & Split=2 \\ \midrule
RTX 3090 latency (ms)                     & 8.15     & 8.91    & 9.48    \\
Orin latency (ms)                         & 28.68    & 29.86   & 33.12 \\ \bottomrule
\end{tabular}
\caption{\textbf{End-to-end Latency:} Unsorted implicit GEMM is up to 1.2$\times$ faster with up to 1.7$\times$ redundant computation.}
\label{tab:ablations:unsorted}
\end{table}

\begin{table}
\centering\small
\begin{tabular}{lccc} \toprule
\textbf{nuScenes-CenterPoint} & Unsorted & Split=1 & Split=2 \\ \midrule
RTX 3090 latency (ms)                     & 3.70     & 3.68    & 3.89    \\
Orin latency (ms)                         & 12.57   & 10.87   & 12.39   \\\midrule
\textbf{Waymo-CenterPoint-1f} & Unsorted & Split=1 & Split=2 \\ \midrule
RTX 3090 latency (ms)                     & 4.60     & 4.33    & 4.49    \\
Orin latency (ms)                         & 19.33    & 17.06 & 19.75 \\ \bottomrule
\end{tabular}
\caption{\textbf{SparseConv Kernel Latency:} Unsorted implicit GEMM kernels could be slower than their mask split counterpart, which is the exact opposite of \tab{tab:ablations:unsorted} results.}
\label{tab:ablations:unsorted_kernel}
\end{table}

As discussed in \sect{sect:kernel_generator}, the design space of \system is a superset of \SpConv v2. We have added several new features to this space, including support for unsorted implicit GEMM, implicit GEMM with an arbitrary number of mask splits ($>2$), and the fetch-on-demand dataflow. The flexibility of \system also allows us to explore different dataflow parameter bindings for \texttt{forward}, \texttt{dgrad}, and \texttt{wgrad} computation. As such, we challenge conventional designs that shares the same dataflow parameters across all kernels. In the following two subsections, we will evaluate the effectiveness of all these new design choices in \system.

\paragraph{Effectiveness of unsorted implicit GEMM.}

\begin{figure}[!t]
    \centering
    \includegraphics[width=\linewidth]{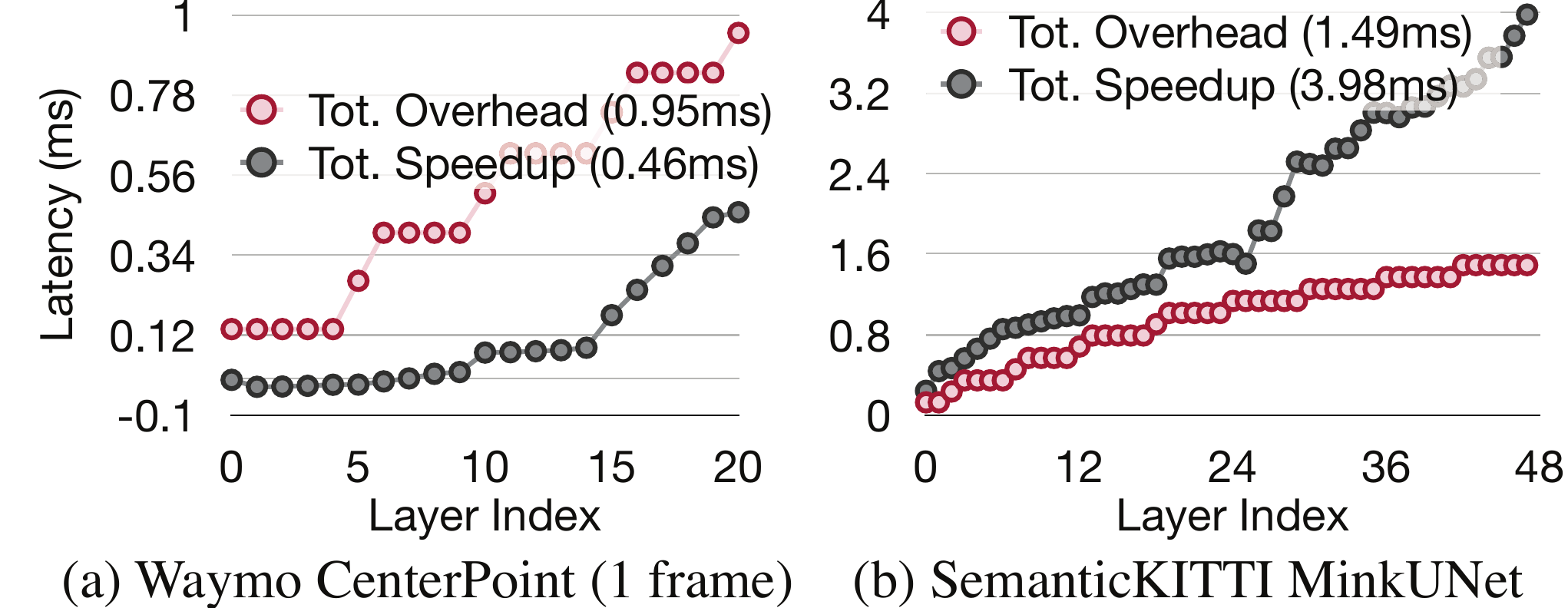}
    \caption{Sorting is able to reduce the computation time, but its overhead outweighs the benefit on detection workloads.}
    \label{fig:ablations:sort_layerwise}
\end{figure}

We first demonstrate the efficacy of unsorted implicit GEMM dataflow (\fig{fig:background:implicit_gemm}) against the sorted implicit GEMM dataflow in \SpConv v2. As in \tab{tab:ablations:unsorted}, the unsorted dataflow is consistently faster on both server and edge GPUs. We further present runtime comparison of all \textit{sparse convolution kernels} between unsorted and sorted dataflows in \tab{tab:ablations:unsorted_kernel}. Interestingly, if we only consider the runtime of convolution kernels, the sorted dataflow is indeed faster. However, the latency difference between \tab{tab:ablations:unsorted} and \tab{tab:ablations:unsorted_kernel} reveals the fact that sparsity-incurred mapping overhead (\eg obtaining the bitmask, sorting the bitmask, performing bitmask reduction and reordering the maps) in the sorted dataflow is non-negligible. %

Moreover, \fig{fig:ablations:sort_layerwise} shows the layerwise comparison of these two versions of \system, in which the gain from reduction in computation is overweighed by the overhead of sorting itself on Waymo object detection. However, sorting does show an advantage on a larger segmentation model (MinkUNet) on the SemanticKITTI benchmark.%
 Our observation challenges the design principle of \SpConv v2, which is to use \textit{amount of computation} as a first-order approximation for end-to-end performance. It also nullifies the assumption that \textit{faster computation kernel} is equivalent to \textit{better end-to-end performance}.

\paragraph{Effectiveness of larger split mask design space.}

\begin{table}
\centering\small
\begin{tabular}{lccc} \toprule
\textbf{Number of splits} & $\{1\}$ & $\{1,2\}$ & $\{0, 1, 2, 3, 4\}$ \\ \midrule
FP16 latency (ms)                    & 16.01     & 15.23    & 14.48    \\
TF32 latency (ms)                         & 26.65   & 23.28   & 21.28   \\
FP32 latency(ms)                         & 35.47    & 29.92 & 25.75 \\ \bottomrule
\end{tabular}
\caption{We evaluated the performance of a SemanticKITTI-MinkUNet workload on an RTX 3090 and found that expanding the design space of implicit GEMM by increasing the number of splits led to up to \textbf{1.4$\times$} improvement compared to the default setting (split=1) in \SpConv v2.}
\label{tab:ablations:more_splits}
\end{table}

We have shown the effectiveness of unsorted implicit GEMM. Additionally, we found that it's also beneficial to have a larger number of splits for segmentation workloads, as demonstrated in \tab{tab:ablations:more_splits}. The parallelism of an implicit GEMM kernel will be increased by $s\times$ with $s$ splits. Because segmentation workloads usually have smaller number of input points, they are more prone to suffer from device under-utilization and increased parallelism will be beneficial. Similarly, the overhead for mapping and partial sum reduction kernels is smaller in segmentation workloads. Significantly reduced computation overhead (\fig{fig:method:num_of_splits}) further supports the preference for a larger number of splits in these scenarios.

\paragraph{Effectiveness of adding fetch-on-demand.}

\begin{figure}[!t]
    \centering
    \includegraphics[width=\linewidth]{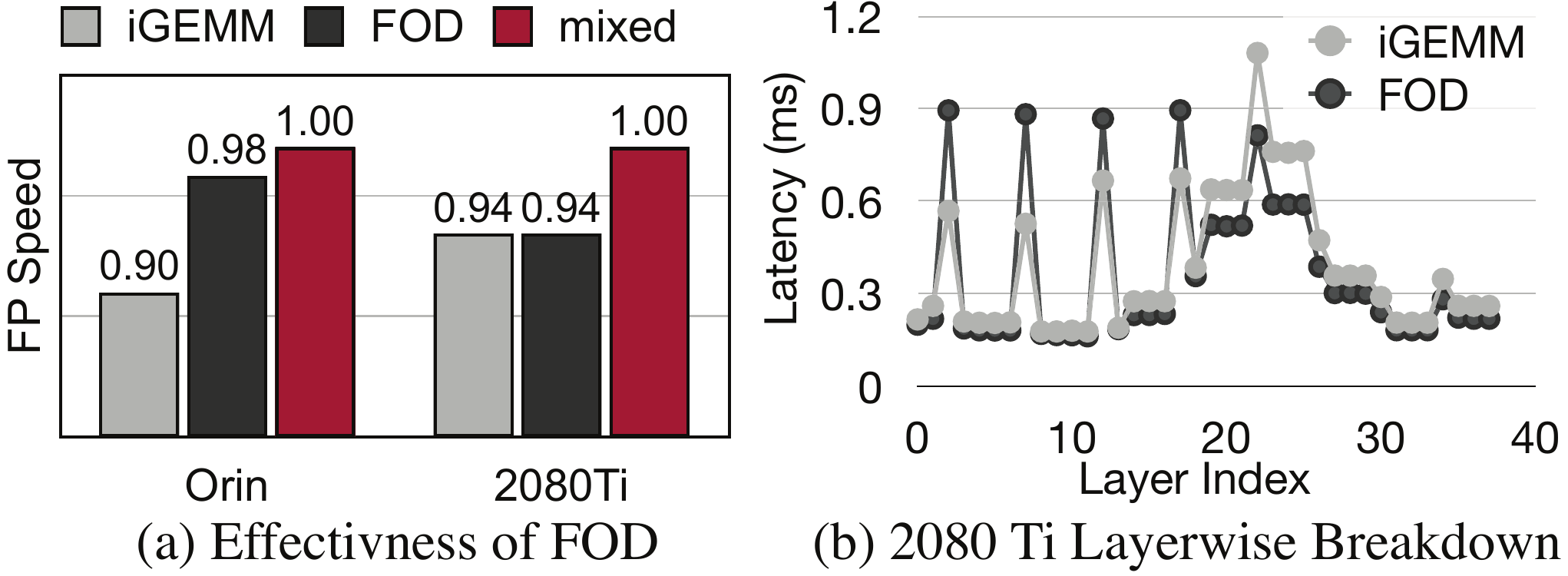}
    \caption{Fetch-on-demand and implicit GEMM dataflows are complementary to each other on FP32 segmentation workloads. A hybrid dataflow is up to 1.06$\times$ faster than the best single dataflow.}
    \label{fig:ablations:igemm_vs_fod}
\end{figure}

We then choose 1-frame MinkUNet on nuScenes running on RTX 2080 Ti and Orin as a benchmark to demonstrate the efficacy of fetch-on-demand dataflow. As in \fig{fig:ablations:igemm_vs_fod}, individually-tuned implicit GEMM and fetch-on-demand dataflows both achieve inferior performance compared with the hybrid dataflow \system. We further present the layerwise latency breakdown of the best tuned implicit GEMM and fetch-on-demand configurations in \fig{fig:ablations:igemm_vs_fod}\textcolor{linkcolor}{b}, where we amortize the mapping time to all layers within each layer group (defined in \sect{sect:autotuner}). The end-to-end performance of fetch-on-demand is notably better than implicit GEMM in decoder layers (\ie layer index $>$ 18) but gets outperformed in downsampling layers, where maps $\mathcal{M}$ could not be reused. This is because implicit GEMM has lower mapping cost while fetch-on-demand computation kernels run faster for the given workload. %

\begin{figure}[!t]
    \centering
    \includegraphics[width=\linewidth]{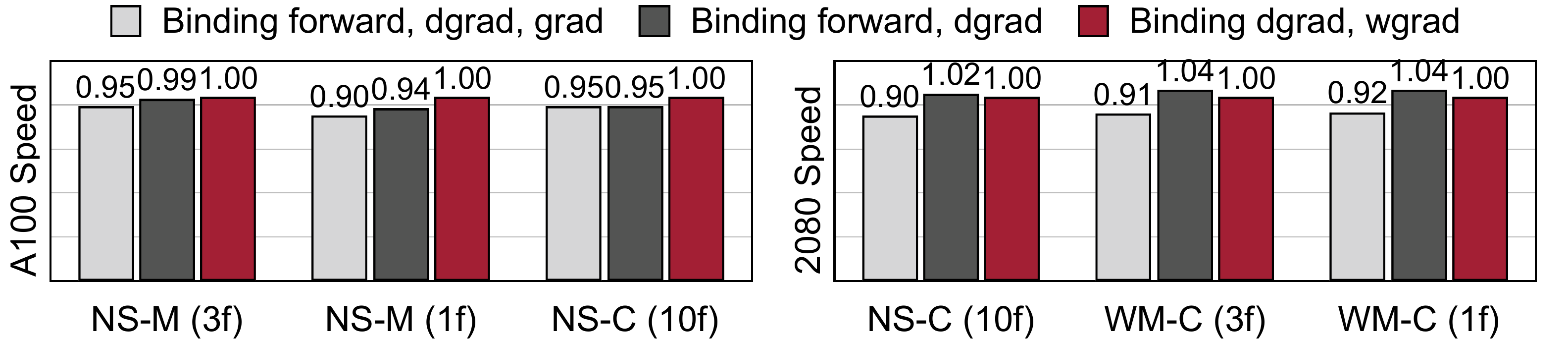}
    \caption{Different from dense kernels, sparse \texttt{forward}, \texttt{dgrad} and \texttt{wgrad} kernels have different preferences for dataflow parameters. Binding hyperparameters for all kernels could hurt the training performance by up to 10\%.}    \label{fig:ablations:bindings}
\end{figure}

\paragraph{Effectiveness of tuner design for training.}
We finally demonstrate that decoupling dataflow parameters for \texttt{forward}, \texttt{dgrad} and \texttt{wgrad} kernels could improve the training performance by up to 10\% in \fig{fig:ablations:bindings}. On both A100 and 2080 Ti, binding parameters for two of the kernels is better than using the same parameters for all three kernels. On A100, binding \texttt{dgrad} and \texttt{wgrad} is better. This is because such strategy could minimize mapping overhead and there is a drastic performance difference (16$\times$) between tensor cores (which runs computation) and CUDA cores (which runs mapping) on A100. On 2080 Ti, binding \texttt{forward} and \texttt{dgrad} is better, since the two kernels share the same workload pattern. Given much smaller performance gap between tensor and CUDA cores on 2080 Ti (3$\times$), the additional mapping overhead for decoupled \texttt{wgrad} and \texttt{dgrad} is acceptable. %

\subsection{Sparse Kernel Generator}
\label{sect:analysis:sparse_kernel_generator}

In this section, we present an analysis of the effectiveness of the design choices outlined in Section \ref{sect:kernel_generator}. Our experiments were conducted on 3090 GPUs using FP32 precision for offline reordering and FP16 precision for all other experiments. Our results demonstrate that simplifying control flows and addressing is critical for achieving optimal performance in sparse kernels. Additionally, we found that the conventional wisdom of fusing GPU kernels as much as possible may not always be applicable in the context of sparse computing.

\begin{figure}[!t]
    \centering
    \includegraphics[width=\linewidth]{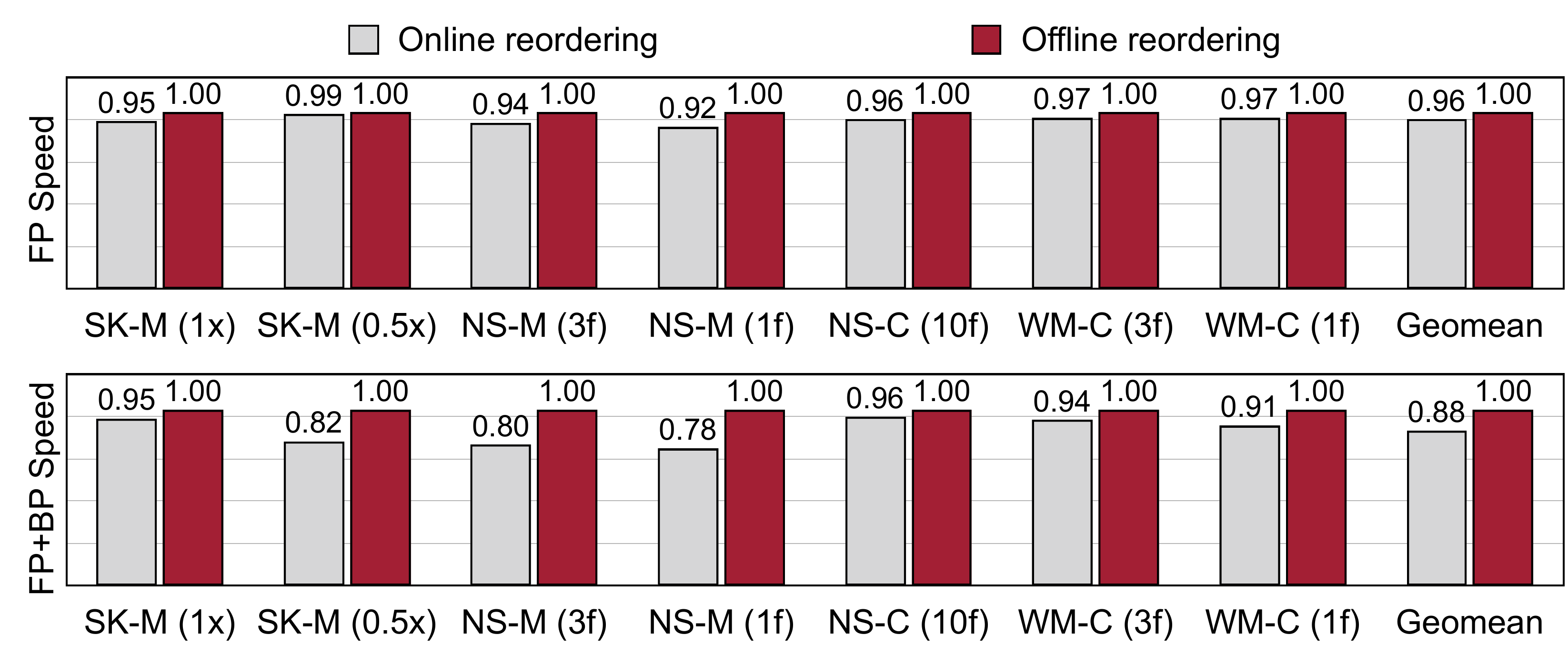}
    \caption{Offline reordering led to a 4\% improvement in inference performance and a 12\% improvement in training performance compared with online reordering.}
    \label{fig:ablations:offline_reordering}
\end{figure}

\begin{figure}[!t]
    \centering
    \includegraphics[width=\linewidth]{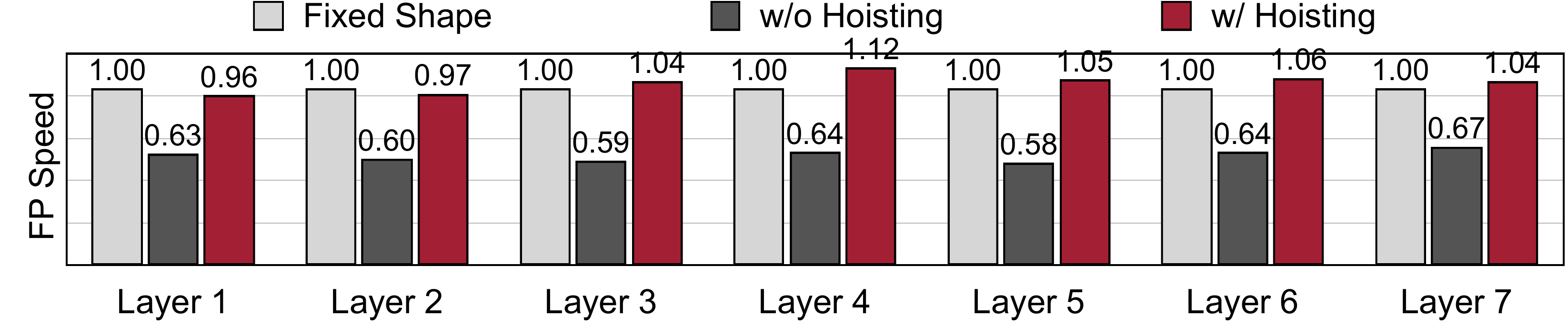}
    \caption{Naively converting fixed shape dense tensor programs to flexible shape sparse convolution kernels will incur 1.5-1.7$\times$ runtime overhead due to repetitive pointer calculation. We bridge such huge performance gap via loop invariance hoisting and show that constant folding is unnecessary for high-performance sparse kernels.}    \label{fig:ablations:hoisting}
\end{figure}

\begin{figure}[!t]
    \centering
    \includegraphics[width=\linewidth]{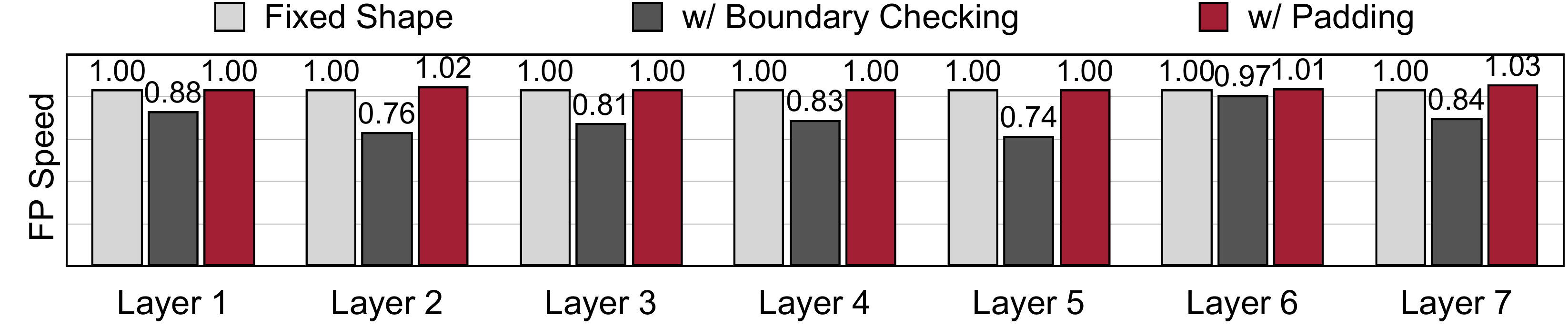}
\caption{Boundary checking when loading \texttt{input\_idx} in \fig{fig:method:sparse_kernel_generator_code} is a major performance bottleneck that leads to 1.3$\times$ latency overhead and can be resolved by offline padding.}
    \label{fig:ablations:boundary_check}
\end{figure}

\begin{figure*}[t]
    \centering
    \includegraphics[width=\linewidth]{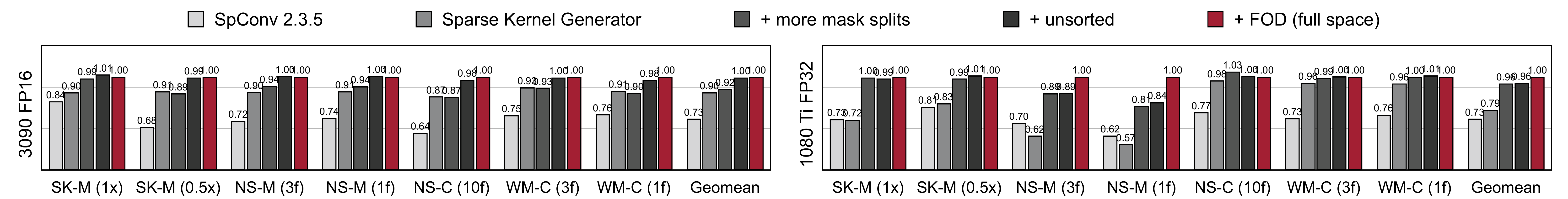}    
    \caption{Summary of performance gain from different techniques and the enlarged design space in \system.}
    \label{fig:analysis:improvement_breakdown}

\end{figure*}

\paragraph{Effectiveness of offline reordering.}
We present the effectiveness of offline reordering in Figure \ref{fig:ablations:offline_reordering}. As described in Section \ref{sect:autotuner}, our approach involves reordering computations based on the values of bitmasks in the implicit GEMM dataflow with mask splitting. While conventional wisdom in GPU kernel design suggests \emph{fusing kernels as much as possible} (including reordering in the sparse convolution kernel), our experiments demonstrate that this can lead to a 4-12\% \emph{reduction} in end-to-end performance compared to offline reordering. Specifically, when considering the \texttt{wgrad} kernels, it is necessary to iterate over the $N_\text{out}$ dimension in the large and innermost $K$ loop. Online reordering 
 introduces an additional level of indirect addressing to the memory access in the innermost loop. This will disrupt the continuous access pattern and results in a significant slowdown for \texttt{wgrad}.

\paragraph{Effectiveness of control flow simplification.} We use MinkUNet on SemanticKITTI as an example to illustrate the importance of simplifying addressing and control flows. In \fig{fig:ablations:hoisting}, we evaluate the benefits of loop invariance hoisting. The results show that a naively converted template can be very inefficient. It is up to 1.7$\times$ slower than the original fixed shape CUDA kernel. However, with the help of loop invariance hoisting in which we move all the common pointer offsets to the outmost possible loop, we can almost totally eliminate the pointer arithmetic overheads. After applying this technique, our templated CUDA kernel can even run slightly faster than the original fixed shape kernels in 5 of 7 sample workloads. \fig{fig:ablations:boundary_check} shows the benefits of reducing control flow instructions by padding the \texttt{map} in \fig{fig:method:sparse_kernel_generator_code}. The instructions performing boundary checking can make the kernel up to 1.3$\times$ slower. Whereas, after eliminating these control flow instructions, this problem can be well solved with the help of padding.

\paragraph{Effectiveness of adaptive tiling.}
We experiment with two sets of tiling sizes in \system dependent upon the MACs of the workload. Adaptive tiling provides up to 1.6$\times$ speedup to \system, compared with fixed tiling version (either always using the small tile sizes or always using the larger tile sizes).

\subsection{Discussions}

\paragraph{Summary on performance gain.} In Figure \ref{fig:analysis:improvement_breakdown}, we present a summary of the performance improvement achieved through the use of our Sparse Kernel Generator and the enlarged design space. Our generator produces high-performance sparse convolution kernels that are $1.1-1.2 \times$ faster than \SpConv 2.3.5, even when using the same dataflow parameters. Remarkably, our code generator comprises only 5\% of the lines of code of \SpConv 2.3.5's metaprogrammer, which significantly reduces system complexity and enhances programmer productivity. For the enlarged design space, more mask splits are very helpful for \textit{segmentation workloads} and \textit{FP32 precision}, while unsorted implicit GEMM is helpful for \textit{detection workloads} and \textit{FP16 precision}. The efficacy of fetch-on-demand is mainly demonstrated in smaller segmentation workloads (\eg \texttt{NS-M}). These results reinforce that there is no one-size-fits-all strategy for sparse kernel design, and that relying on first-order approximations for end-to-end performance is unreliable.

\paragraph{Insights for microarchitectural improvements.} Our \system also provides new insights for future microachitecture design. Our findings indicate that when memory bandwidth is halved on RTX 3090, the latency of the system increases by \textbf{1.2$\times$}. In contrast, reducing peak computation throughput by a factor of 2 results in a more substantial slowdown of \textbf{1.4$\times$}. Therefore, scaling computation units instead of off-chip memory bandwidth can provide more effective improvements. Moreover, it is apparent from \tab{tab:ablations:unsorted} and \tab{tab:ablations:unsorted_kernel} that mapping operations account for up to 50\% of the total runtime. Leveraging the efficient ASIC design~\cite{lin2021pointacc} for these operators could significantly enhance the performance of GPUs when executing sparse computation workloads.%

\paragraph{Future applications.} \system platform presents novel opportunities for enhancing machine learning workloads beyond point clouds and graphs. For instance, in image segmentation~\cite{li2017not} and video recognition ~\cite{pan2018recurrent}, not all pixels hold equal significance. Hence, the selective computation on a sparse subset of pixels using \system can potentially significantly enhance efficiency. Furthermore, masked autoencoders (MAEs)~\cite{he2021masked} exhibit inherent sparsity in input patterns during training. While existing approaches already attempt to exploit this sparsity using sparse convolution~\cite{huang2022green, tian2023designing}, we posit that \system has the potential to unlock even greater speedups for such workloads.

\section{Related Work}

\paragraph{Compiler-Based Tensor Program Optimization.} Our system benefits from recent advances in tensor program compilation. The pioneering research TVM \cite{chen2018tvm} provides graph-level and operator-level abstractions for deep learning workloads based on the essence of Halide~\cite{ragankelley2013halide}. Based on TVM, AutoTVM \cite{chen2018learning} automatically discovers the optimal mapping of a fixed-shape tensor program onto the target hardware. Nimble~\cite{shen2021nimble} and DietCode~\cite{zheng2022dietcode} are compilers stemmed from TVM that can generate tensor programs with dynamic-shape workloads, but they are still tailored for dense workloads (\eg transformers with variable length input sequences) and cannot deal with the sparsity in point clouds. More recently, TensorIR \cite{feng2022tensorir} proposed a new IR for tensor programs and allows easier tensorization of accelerator primitives. SparseTIR~\cite{ye2022sparsetir} further extended TensorIR to support sparse workloads. Bolt~\cite{xing2022bolt} combines the advantages of fully-automatically generated kernels~\cite{chen2018tvm} with hand-written subroutines~\cite{nvidia2022cutlass} through graph matching.

\paragraph{Point Cloud Accelerators.}
Deep learning on point clouds has also generated considerable interest in domain-specific accelerator design.  Zhu~\etal~\cite{zhu2020spcnn} proposed a sparsewise dataflow that skips cycles for zero-weight computations and saves energy through gating. Mesorasi~\cite{feng2020mesorasi} co-designed its architecture with the delayed aggregation algorithm to reduce redundant computation in point cloud NNs. More recently, Point-X~\cite{zhang2021pointX} exploited spatial locality in point clouds through clustering, mapping point clouds into distributed computation tiles. It maximized parallelism and minimized data movement. Additionally, PointAcc~\cite{lin2021pointacc} mapped all mapping operators in point cloud NNs to a versatile bitonic sorter, making it the first specialized accelerator to support 3D sparse convolution computation. Crescent~\cite{feng2022crescent} tamed irregularities in point clouds through approximate neighbor search and selectively elided bank conflicts, while Ying et al.~\cite{ying2022pushing} pushed point cloud compression to edge devices through intra- and inter-frame compression.

\section{Conclusion}

We introduce \system, a high-performance GPU sparse computation library designed for point cloud and graph deep learning. \system features a highly optimized Sparse Kernel Generator with less than one-tenth of the engineering cost compared with the state-of-the-art system. It further enables us to build an input-aware Sparse Autotuner that selects the best configuration for each layer. \system achieves \textbf{1.7-3.3$\times$} inference speedup and \textbf{1.2-3.7$\times$} faster training compared to state-of-the-art \ME, \SpConv v1/v2, and \systemv1 on seven real-world perception workloads. \system also achieves \textbf{2.6-7.6$\times$} speedup over DGL, PyG and Graphiler when running R-GCNs. We hope that \system will facilitate future system and microarchitectural research in sparse computation on 3D data and graphs.

\begin{acks} 
We would like to thank Yan Yan and Bohan Hou for helpful discussions. This work was supported by MIT-IBM Watson AI Lab, MIT AI Hardware Program, MIT-Amazon Science Hub, NSF and Hyundai Motor. Zhijian Liu was partially supported by the Qualcomm Innovation Fellowship.
\end{acks}

\bibliographystyle{ACM-Reference-Format}

\end{document}